\documentclass[twocolumn,twocolappendix]{aastex631}

\usepackage{amsmath}
\usepackage{empheq}
\usepackage{graphicx}
\expandafter\let\csname equation*\endcsname\relax
  \expandafter\let\csname endequation*\endcsname\relax 
\usepackage{mathrsfs}
\usepackage{textcomp}
\usepackage{enumitem}   
\usepackage{gensymb}
\usepackage{hyperref}
\usepackage{graphicx}
\usepackage[caption=false]{subfig}
\usepackage{multirow}
\usepackage{longtable}
\usepackage[flushleft]{threeparttable}
\usepackage{booktabs} 
\usepackage{CJK}
\usepackage{hyperref}
\usepackage{multirow}
\hypersetup{colorlinks, linkcolor={blue}, citecolor={blue}, urlcolor={blue}} 

\usepackage[mathlines]{lineno}

\definecolor{DarkOrange}{RGB}{204, 85, 0}
\definecolor{LincolnGreen}{RGB}{17, 102, 0}

\usepackage{xspace}

\newcommand\swift{\textit{Swift}\xspace}
\newcommand\xmm{\textit{XMM-Newton}\xspace}

\newcommand\rosat{\textit{ROSAT}\xspace}

\newcommand\hst{\textit{HST}\xspace}
\newcommand\hstlong{\textit{Hubble Space telescope}\xspace}
\newcommand\swiflong{\textit{Neil Gehrels Swift Observatory}\xspace}

\newcommand\fluxdens{erg cm$^{-2}$ s$^{-1}$ {$\rm\AA^{-1}$}\xspace}

\newcommand\Rin{$R_{\rm in}$\xspace}
\newcommand\Rins{$R_{\rm in}^{*}$\xspace}
\newcommand\Tp{$T_{\rm p}$\xspace}
\newcommand\Rratio{$R_{\rm out}/R_{\rm in}$\xspace}

\newcommand\Rout{$R_{\rm out}$\xspace}
\newcommand\Rorb{$R_{\rm orb}$\xspace}

\newcommand\Mbh{$M_{ \rm BH}$\xspace}
\newcommand\ergs{erg s$^{-1}$\xspace}
\newcommand\msun{$M_\odot$\xspace}

\newcommand\target{GSN\,069\xspace}

\shorttitle{GSN\,069 X-ray + UV}
\shortauthors{Guolo et al.}

\begin{document}
\pagenumbering{arabic}

\title{The properties of GSN\,069 accretion disk from a joint X-ray and UV spectral analysis: \\ stress-testing quasi-periodic eruption models
}

\author[0000-0002-5063-0751]{M. Guolo}
\affiliation{Bloomberg Center for Physics and Astronomy, Johns Hopkins University, 3400 N. Charles St., Baltimore, MD 21218, USA}
\correspondingauthor{Muryel Guolo}
\email{mguolop1@jhu.edu}

\author{A. Mummery}
\affiliation{Oxford Theoretical Physics, Beecroft Building, Clarendon Laboratory, Parks Road, Oxford, OX1 3PU, United Kingdom}

\author[0000-0002-4043-9400]{T. Wevers}
\affiliation{Astrophysics \& Space Institute, Schmidt Sciences, New York, NY 10011, USA}
\affiliation{Space Telescope Science Institute, 3700 San Martin Drive, Baltimore, MD 21218, USA}

\author[0000-0002-2555-3192]{M. Nicholl}
\affiliation{Astrophysics Research Centre, School of Mathematics and Physics, Queens University Belfast, Belfast BT7 1NN, UK}

\author[0000-0003-3703-5154]{S. Gezari}
\affiliation{Space Telescope Science Institute, 3700 San Martin Drive, Baltimore, MD 21218, USA}
\affiliation{Bloomberg Center for Physics and Astronomy, Johns Hopkins University, 3400 N. Charles St., Baltimore, MD 21218, USA}

\author[0000-0002-5311-9078]{A. Ingram}
\affiliation{School of Mathematics, Statistics, and Physics, Newcastle University, Newcastle upon Tyne, NE1 7RU, UK}

\author[0000-0003-1386-7861]{D. R. Pasham}
\affiliation{MIT Kavli Institute for Astrophysics and Space Research, 70 Vassar Street, Cambridge, MA 02139, USA}

\begin{abstract}

We present an analysis of \textit{Hubble Space Telescope (HST)} and \textit{XMM-Newton} data of the tidal disruption event (TDE) candidate and quasi-periodic eruption (QPE) source GSN\,069. Using ultraviolet (UV) and optical images at \textit{HST} resolution, we show that GSN\,069’s emission consists of a point source superimposed on a diffuse stellar component. The latter accounts for $\leq 5\%$ of the UV emission in the inner 0.5"$\times$0.5" region, while the luminosity of the former cannot be attributed to stars. Analyzing the 2014/2018 \hst UV spectra, we show that to leading order the intrinsic spectral shape is $\nu\,L_{\nu}\propto\nu^{4/3}$, with $\sim10\%$ far UV flux variability between epochs. The contemporaneous X-ray and UV spectra can be modeled self-consistently in a thin disk framework. At observed epochs, the disk had an outer radius ($R_{\rm out}$) of $\mathcal{O}(10^3R_{\rm g})$, showing both cooling and expansion over four years. Incorporating relativistic effects via numerical ray tracing, we constrain the disk inclination angle ($i$) to be $30^\circ\,\lesssim\,i\,\lesssim\,65^\circ$ and identify a narrow region of spin-inclination parameter space that describes the observations. These findings confirm that GSN\,069 hosts a compact, viscously expanding accretion disk likely formed after a TDE. Implications for QPE models are: (i) No published disk instability model can explain the disk’s stability in 2014 (no QPEs) and its instability in 2018 (QPEs present); (ii) While the disk size in 2018 allows for orbiter/disk interactions to produce QPEs, in 2014 the disk was already sufficiently extended, yet no QPEs were present. These findings pose challenges to existing QPE models.


\end{abstract}

\vspace{1em}

\section{Introduction}

X-ray emission from \target was first detected in July 2010 during an \xmm slew survey \citep{Saxton2011} observation, with a flux hundreds of times higher than previous upper limits from \rosat observations over a decade earlier. Monitoring with the \swiflong (hereafter \swift) and \xmm revealed a gradual decline in flux over subsequent years \citep{Miniutti2013}. The extremely soft X-ray spectrum, which cools as the source evolves \citep[][see their Figure 4 for a long-term X-ray light curve]{Miniutti2023_noqpe}, seems to be consistent with a viscously evolving accretion disk formed after a tidal disruption event \citep[TDE,][]{Cannizzo1990,Mummery2020}. The source also exhibits luminosities and temperatures similar to those of X-ray bright TDEs, including both optically and X-ray discovered sources \citep{Saxton2020,Mummery2023,Guolo2024b}.

\xmm observations of \target in late 2018 and early 2019, revealed high-amplitude X-ray flares, increasing up to two orders of magnitude in the hardest energy bands. These short-lived flares, occurring roughly every nine hours, marked the first detection of quasi-periodic eruptions \citep[QPEs,][]{Miniutti2019}. QPEs are now a well established phenomena in high-energy astrophysics, with seven confirmed sources 
\citep{Giustini2020,Arcodia2021,Arcodia2024,Nicholl2024} and few candidates \citep{Evans2023,Guolo2024,Chakraborty2021,Quintin2023}. These eruptions are marked by thermal-like X-ray spectra with typical temperatures of 100–200 eV, superimposed on more stable, cooler disk emission.

Between late 2019 and 2022, \target's quiescent disk emission re-brightened, the origin of which is still unknown. One possibility is that the surviving core of the partially disrupted star returned, leading to increased accretion as new mass deposition in the disk is needed. This rebrightening coincided with the disappearance of QPEs \citep{Miniutti2023_noqpe}.

The origin of QPEs remains a topic of debate. One possibility is that they arise from instabilities within the accretion disk itself \citep{2022ApJ...928L..18P, Pan2023, Sniegowska2023, Kaur2023}. Alternatively, QPEs could result from recurring interactions between orbiting bodies (such as a star or compact object) and the central massive black hole (MBH). The reader is refereed to  \citet{Zhao2022, King2022,lu23,Linial_Sari2023, Linial2023,Franchini2023} for some of the many available models. In these orbit related models, QPE would be an electromagnetic counter-part of a extreme mass ratio inspirals (EMRI) system.

The orbiter/disk collision models \cite[e.g.,][]{Linial2023, Franchini2023} have gained traction by reproducing key features of QPE evolution, such as alternating short and long recurrence intervals and counterclockwise ``hysteresis loops" in temperature (hardness) and luminosity during eruptions \citep[e.g.,][]{Arcodia2022}, which are naturally explained by this model \citep{Vurm2024}. The model also predicts a direct link between QPEs and TDEs, a connection now unambiguously confirmed by \citet{Nicholl2024}.

\begin{figure*}[t!]
	\centering
	\includegraphics[width=0.9\columnwidth]{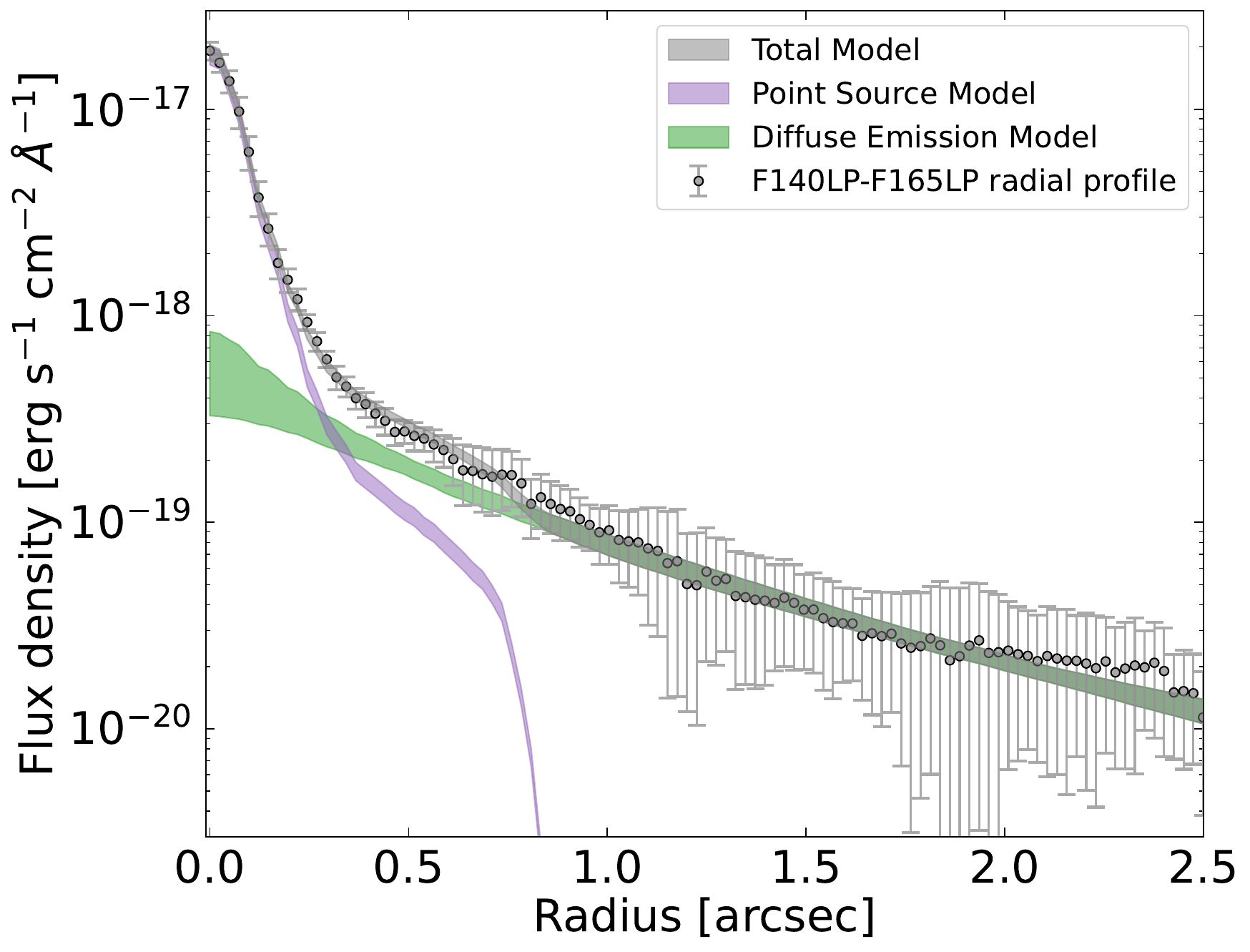} \includegraphics[width=0.9\columnwidth]{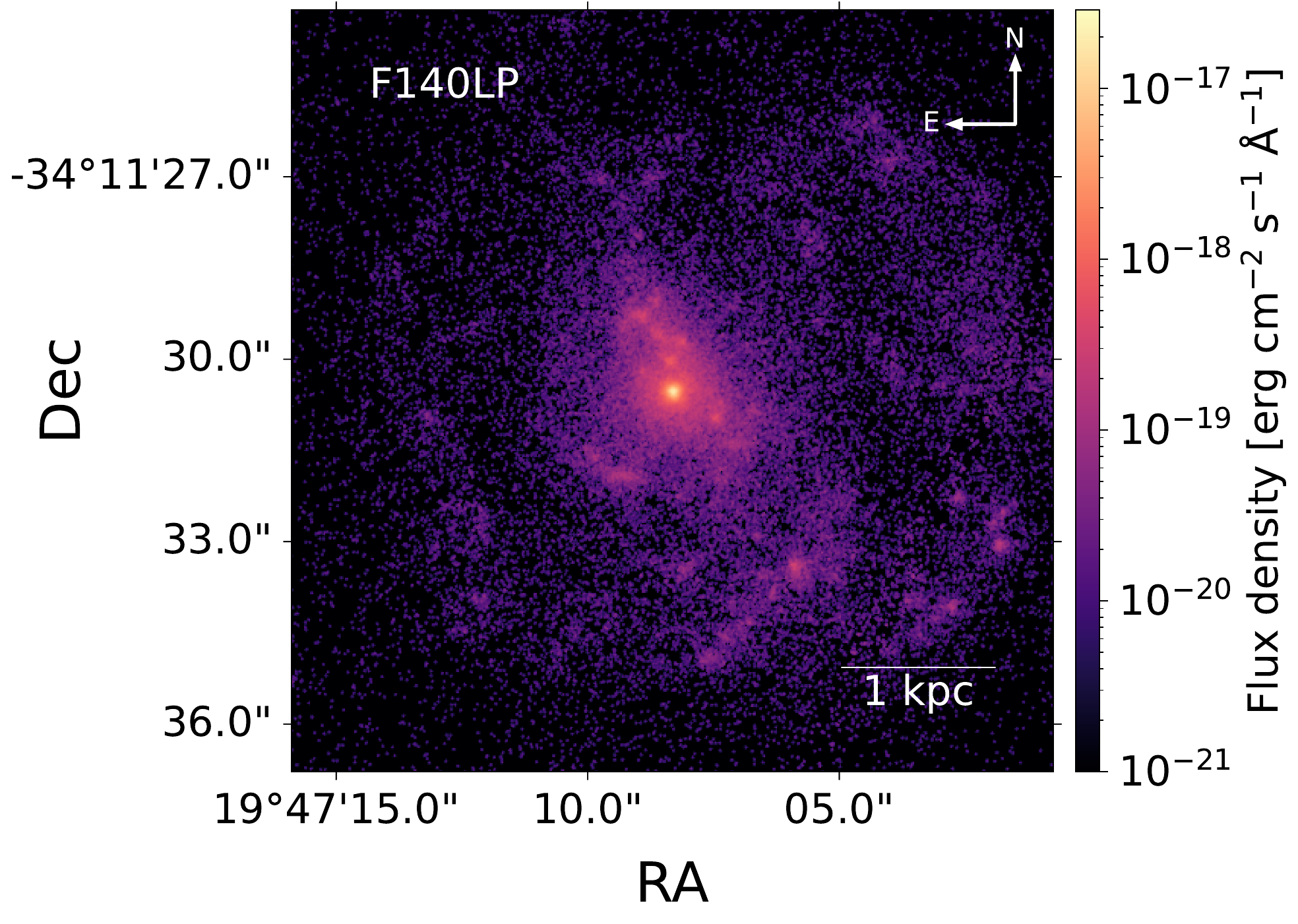}\\ 
    \vspace{0.3cm}
    \includegraphics[width=0.9\columnwidth]{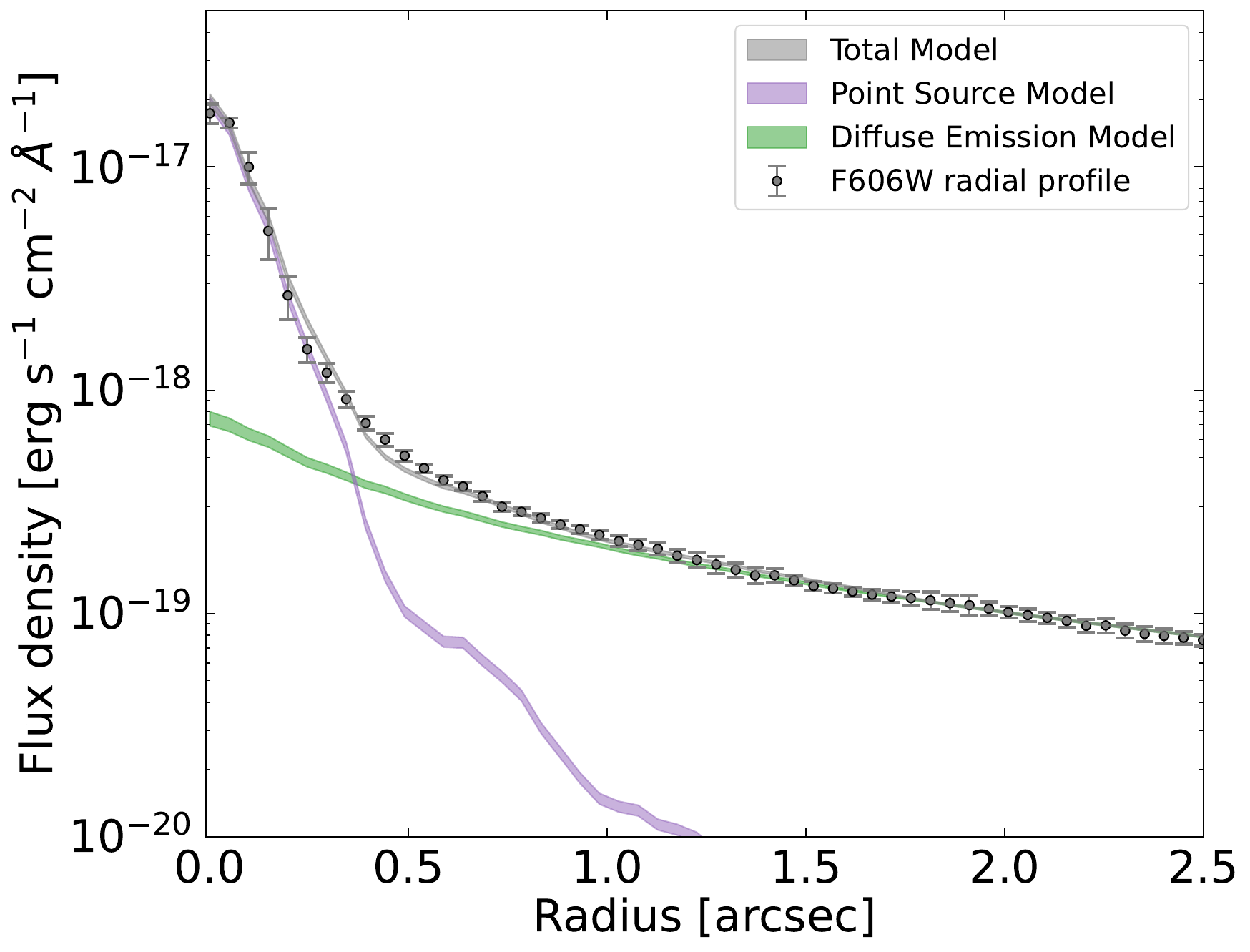} \includegraphics[width=0.9\columnwidth]{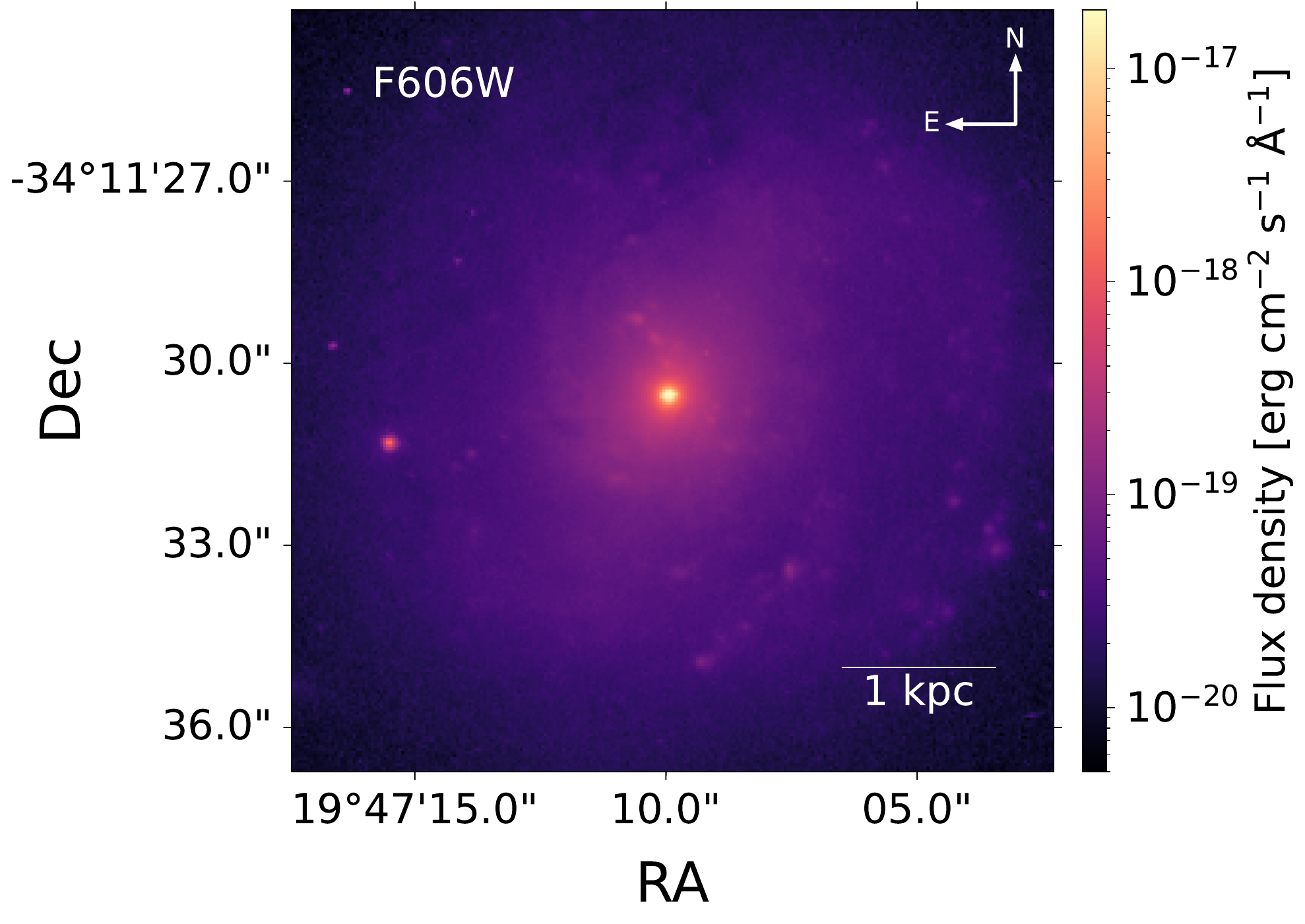}
 
	\caption{{\bf Left:} Radial profile fitting, for F6060W filter (Top) and red-leak corrected FUV filter (F140LP-F165LP, Bottom). Data is shown in gray points, total model in gray band, point source model in purple, and diffuse galaxy emission (Sersic) profile in green. All bands are 68\% of the model's posteriors. The point source dominates the inner pixels. {\bf Right:} F6060W (Top) and F140LP (Bottom) \hst/SBC images of \target. Color scheme shows flux density (\fluxdens) in logarithm scale. }
    \label{fig:profile_fit}
\end{figure*}

However, for these models to hold, the orbital radius (\Rorb) of the EMRI must be smaller than the outer disk edge (\Rout), i.e., \Rorb $\leq$ \Rout for all QPE sources. While \Rorb depends solely on the QPE timing properties and the black hole mass (\Mbh), estimates of \Rout are more challenging, as the X-ray quiescent spectrum probes only the inner disk properties. Estimates on \Rout can however be obtained through a self-consistent analysis of contemporaneous X-ray, ultraviolet (UV), and/or optical data \citep[e.g.,][]{Mummery2020,Nicholl2024,mummery2024fitted,Guolo_Mummery2024}.

High-quality UV/optical data from the \hst is available for \target and has been presented in a few studies \citep{Miniutti2019,Sheng2021,Miniutti2023_longterm}. However, these analyses diverge in their interpretations, particularly regarding the fraction of UV emission produced by stars versus accretion. Notably, the extent to which the observed UV emission originates from the quiescent disk (observed in X-rays) remains to be fully explored, and a self-consistent analysis of the X-ray and UV spectra is yet to be conducted.

In this paper, we present a detailed analysis of UV/optical photometric and spectroscopic data, along with a self-consistent, simultaneous fitting of the UV and X-ray spectra. Our aim is to determine the nature of the UV/optical emission in \target and constrain key physical parameters through full spectral energy distribution (SED) modeling. In future work (Guolo \& Mummery, in prep), we will present a fully time-dependent solution for \target's disk evolution, providing insights into its `long-lived TDE' nature. 

The structure of this paper is as follows: in \S\ref{sec:data}, we describe the UV and optical data analysis and including UV/optical photometry (\S\ref{sec:image}) and UV spectroscopy (\S\ref{sec:spec}). While joint X-ray and UV spectra fitting is performed in \S\ref{sec:SED}. In \S\ref{sec:discus}, we discuss the implications for the nature of \target and constraints on QPE models. Our conclusions are presented in \S\ref{sec:conclusions}. 

We adopt a standard $\Lambda$CDM cosmology with a Hubble constant $H_0=73\,{\rm km\,s^{-1}\,Mpc^{-1}}$ \citep{Riess2022}. A Bayesian statistics framework is considered throughout the paper, we denote the marginalized posterior of a parameter $\theta$ of a given model as $P(\theta | \rm{data})$, and the prior as $\pi(\theta)$. Inferred parameters are reported as median of the marginalized posterior, and the uncertainties correspond to the bounds that contains  68\% credible intervals, unless otherwise stated. Model contours are also 68\% of their posteriors, unless otherwise stated. These differ from the frequentist definition historically used in X-ray studies \citep[see][for relevant discussion]{Andrae2010,Buchner2014,Buchner2023}.

\section{Optical and UV Data Analysis} \label{sec:data}
This work is entirely based on publicly available \hstlong (\hst) and \xmm data. It includes \hst/SBC images taken between August and November 2020 in three filters F606W, FL140LP and F165LP. Two epochs of \hst/STIS far (G140L) and near (G230L) UV spectra, taken in 2014 adn 2018. Two epochs of \xmm EPIC-pn X-ray spectra, taken contemporaneous to the two UV spectra. The detailed reduction of the data, including the removal of the QPE emission from the 2018 X-ray spectrum, is presented in \S\ref{app:reduction}. No QPEs were present in the 2014 X-ray observation of \target. In Table \ref{tab:data} a summary of the data is presented. For context, if the initial event (e.g., a TDE) that triggered \target's emission occurred shortly before its discovery in 2010 \citep{Saxton2011}, the spectral observations examined in this study would correspond to approximately 4 and 8 years after the event.

\begin{deluxetable*}{ccccc}
\label{tab:data}
\tablecaption{Data Information.}
\tabletypesize{\small}
\tablehead{
\colhead{Data Type} & \colhead{Observatory/instrument} & \colhead{Filter/Grating} & \colhead{Wavelength Coverage} & \colhead{Obs Date (UT)}
}
\startdata
X-ray Spectra & \xmm\ EPIC-pn & Thin & 1.2--41.3~{\AA} (0.3--10 keV) & 2014-12-05, 2018-12-24 \\ \hline
UV Spectra & \hst\ STIS & G140L + G230L & 1150--3150~{\AA} & 2014-12-14, 2018-12-31 \\\hline
UV and Optical Imaging & \hst\ SBC & F140LP & $\lambda_c = 1476~{\AA}$, FWHM = 226~{\AA} & 2020-11-19 \\
 & & F165LP & $\lambda_c = 1740~{\AA}$, FWHM = 200~{\AA} & 2020-11-19 \\
 & & F606W & $\lambda_c = 5962~{\AA}$, FWHM = 2253~{\AA} & 2020-08-16 \\
\enddata
\end{deluxetable*}

\subsection{\hst Optical and Far UV Images}\label{sec:image}

The nature of the UV emission in \target has been interpreted differently in previous studies. \citet{Miniutti2019} argued that the UV emission was dominated by young stars, citing a lack of variability between two epochs of UV spectroscopy and a visual comparison with B-type star spectra. In contrast, \citet{Sheng2021} used a spectral decomposition method, assuming linear combination of a power-law spectrum, $F_{\lambda} \propto \lambda^{-\alpha}$ (with $\alpha$ as a free parameter), and three B-type star spectra with distinct temperatures. They found that about 80\% of the UV emission came from the power-law component. Similar to \citet{Miniutti2019}, they also reported no UV spectral variability between 2014 and 2018. In this section, we re-examine this issue using spatial decomposition and flux/color data from multi-band \hst images. The question on the variability will be addressed in \S\ref{sec:spec}, as only one epoch of imaging in each filter is available.

In the right panels of Fig.~\ref{fig:profile_fit}, we show the F606W and F140LP images. The extended stellar continuum of the host galaxy is clear in F606W, the F140LP is however more concentrated and `blobby', tracing the star formation regions around the galaxies, which seems to be concentrated in the circumnuclear region as well as spread throughout the spiral arms of the galaxy. A very bright nucleus is clear in both filters (note the logarithm scale), and its nature can be elucidated by determining whether the emission is spatially resolved or point-like.

Following previous work \cite[e.g.,][]{French2020,Patra2023}, we decompose the radial profile of the images with a combination of a Sersic profile and a point source, such that the observed surface flux density of a given filter ($\mu_{o,\lambda}$) can be described by the follow convolution operation:

\begin{equation}\label{eq:1}
    \mu_{o,\lambda}(r) = \int \mu_{i,\lambda}(r') K_\lambda(r - r') {\rm d}r'
\end{equation}

\noindent where  $K_\lambda(r - r')$ is the point spread function (PSF)\footnote{We use the standard PSF models for the filters as available in the \href{https://www.stsci.edu/hst/instrumentation/wfc3/data-analysis/psf}{HST Documentation}.} of the filter of wavelength $\lambda$, and $\mu_{i,\lambda}$ is the intrinsic model for the total emission, and written as:

\begin{equation}\label{eq:2}
    \mu_{i,\lambda}(r) = A_\lambda\delta(r) + \mu_{S,\lambda}\,{\rm exp}(-k\,r^{1/n}) 
\end{equation}

\noindent  where $\delta(r)$ is a Dirac delta function, $A$ is the total flux of the point source, and right hand term is the Sersic profile of index $n$. The HST/SBC UV filters have a known problem of red leak for wavelengths $\geq2000$\AA. To remove the contaminating flux from longer wavelengths, and to get the true FUV flux, we subtracted the two far-UV images F140LP and F165LP, as recommended by the ACS instrument handbook
\footnote{\href{https://hst-docs.stsci.edu/acsihb/chapter-4-detector-performance/4-4-the-sbc-mama}{ACS handbook}}.
We then use \texttt{ultranest} \citep{Buchner2019} to fit the radial profile model (Eq.~\ref{eq:1} and \ref{eq:2}) to both F606W and the red-leak corrected FUV images (F140LP-F165LP). The results of the fits are shown in the left panels of Fig.~\ref{fig:profile_fit}.

\begin{figure}[h!]
	\centering
	\includegraphics[width=\columnwidth]{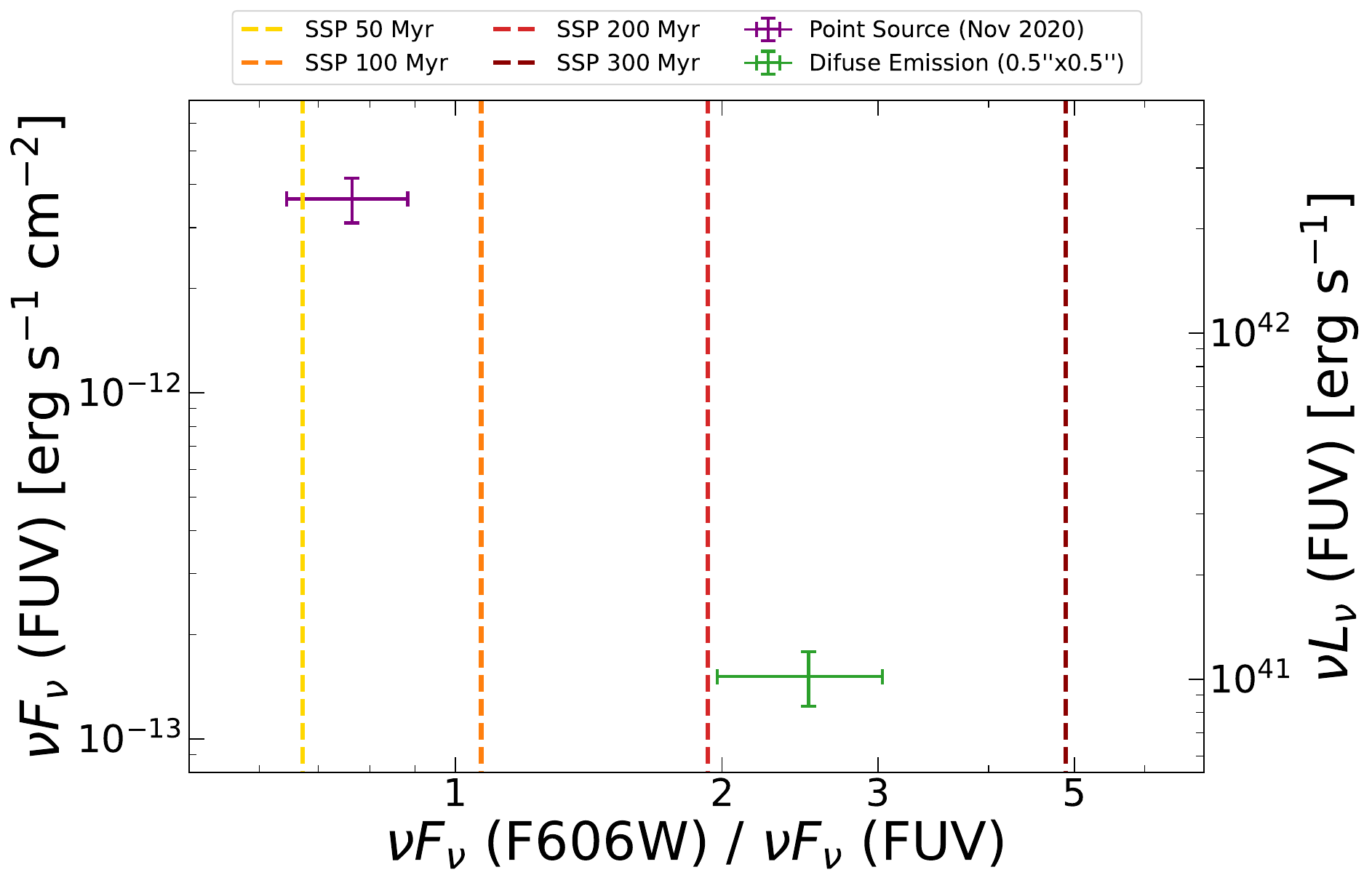}
 
	\caption{Color vs. flux/luminosity \target's UV/optical emission. Purple refers to the point-source (size $\leq$ 0.1\arcsec or 35 pc), while green is the diffuse emission as measured in a 0.5\arcsec$\times$0.5\arcsec aperture. Flux/luminosities and colors are intrinsic, i.e., corrected for both Galactic and host-galaxy extinction. Point-source is too bright to be power by stellar emission (see text for details). Error-bars are mostly dominated by the uncertainty on the intrinsic E(B-V). The vertical lines show the color of simple stellar populations of distinct ages.}
    \label{fig:point_source}
\end{figure}

\begin{figure*}[t!]
	\centering
	\includegraphics[width=\textwidth]{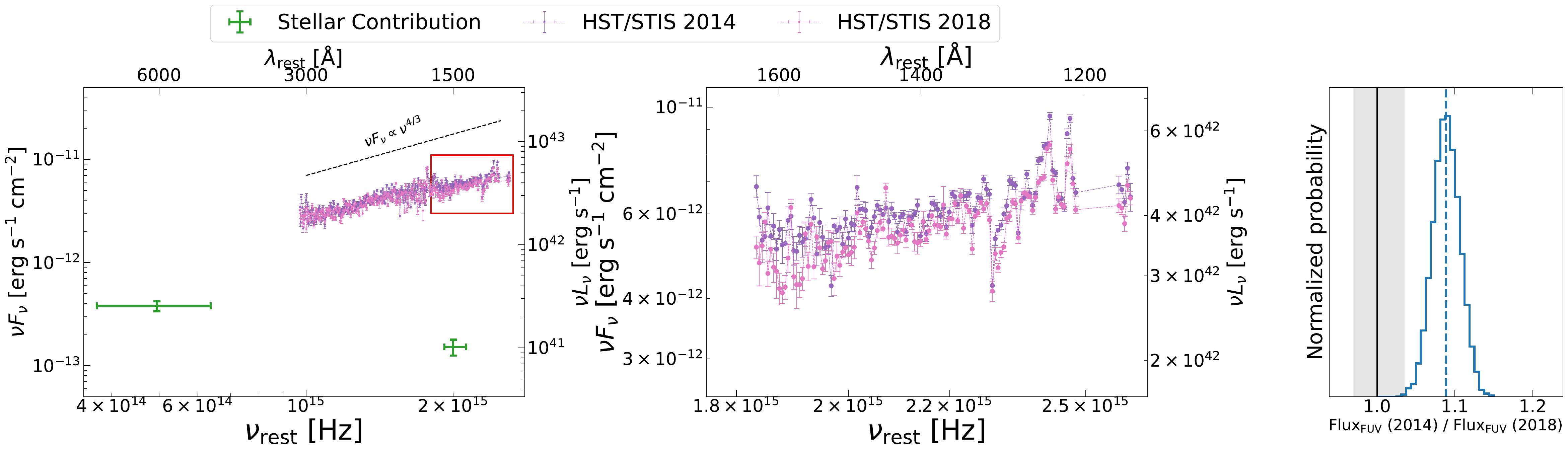}
 
	\caption{{\bf Left:} extinction-corrected \hst/STIS UV spectra of \target in 2014 (purple) and 2018 (pink). Green point show estimate of the stellar contribution to the inner aperture that the spectra were extracted. {\bf Middle:} Zoom into Far UV (FUV) portion of the spectra.  
      {\bf Right:} The probability distribution from $10^4$ simulations of the integrated FUV flux ratio (Eq. \ref{eq:2}) between the two epochs. The gray band represents the maximum 3\% systematic instrumental variability expected from STIS. A $\sim$10\% decrease in flux between epochs is statistically significant.}
    \label{fig:spec}
\end{figure*}

The fitting results are clear from the left panel of Fig.~\ref{fig:profile_fit}: in both filters, GSN-069’s photometry is well-described by a combination of diffuse emission and a point source, with the latter dominating the flux in the central pixels by a factor $\gtrsim$ 20. Since the optical/UV emission is unresolved, its physical size is smaller than the full width at half maximum (FWHM) of PSF of the filters, which, for SBC’s filters, is approximately $0.1\arcsec$. Given \target’s luminosity distance of 73 Mpc ($z \sim 0.018$), this translates to a maximum physical size of 35 pc. Such compact emission could either originate from a nuclear stellar cluster (NSC) or be due to accretion onto the MBH. The flux/luminosity and color of the emission can then be used to distinguish between these two scenarios.

To convert from observed fluxes to the intrinsic luminosity, corrections for both dust Galactic extinction and intrinsic attenuation in the host galaxy need to be applied. For the Galactic extinction we use color-excess of $E(B-V)_{\rm G} = 0.023$ \citep{Schlafly2011} and \citet{Cardelli1989} extinction law. For the intrinsic attenuation we use \citet{Calzetti2000} law and $E(B-V) = 0.10 \pm 0.01$, as derived from the Balmer decrement of the central resolution element's spectrum obtained from the MUSE integral field Unit (IFU) data of \target as presented and analyzed in \citet{Wevers2024}. We will show in \S\ref{sec:SED} that this $E(B-V)$ is consistent with the one independently measured from the full SED fitting. The intrinsic FUV luminosity and the F606W/FUV color of the point source are shown Fig.~\ref{fig:point_source}.

A nuclear starburst could, in principle, produce a UV-bright, centrally concentrated population of young stars. To evaluate whether a nuclear star cluster (NSC) could account for the observations, we utilize simple stellar population (SSP) models from \citet{Maraston05} to reproduce the FUV luminosity. For a stellar population age of 100 (50) Myr—consistent with its F606W/FUV color (Fig.~\ref{fig:point_source})—reproducing the UV luminosity ($L_{\rm FUV} \approx 2 \times 10^{42}$ erg s$^{-1}$) requires a stellar mass (young stars in the NSC) of log($M_{\rm NSC}$) $\sim$ 7.8 (8.2) $M_{\odot}$. These mass estimates alone would classify it among the most massive known NSCs \citep{Seth06, Neumayer20}. Considering that young stellar populations typically account small fraction the total NSC mass, the total mass would be even greater.
However, even more constraining than the mass of the young stars is the resulting star formation rate surface density ($\Sigma_{\rm SFR}$). Given its point source nature, the region can span no more than $\lesssim 35$ pc. Assuming the stars formed over a timescale similar to their current age, $\Sigma_{\rm SFR}$ would be on the order of $\mathcal{O}(100 \, M_{\odot}  \text{yr}^{-1}  \,   \text{kpc}^{-2})$. This is more than two orders of magnitude higher than the most intense star-forming regions observed in all disk/spiral galaxies of the SDSS MaNGA survey \citep[$\sim 1 M_{\odot}  \ \text{yr}^{-1} \ \text{kpc}^{-2}$,][]{Law2022}.
We discuss the $\Sigma_{\rm SFR}$ computation in detail in Appendix \ref{app:sfrsd}.

Thus, attributing (any relevant fraction of) the UV-bright point source to an NSC is highly contrived, as it would mean \target's NSC would have, by far, the highest $\Sigma_{\rm SFR}$ in the local universe. As we will show below, emission from an accretion disk provides a more natural and self-consistent explanation for the full SED. We also compute the luminosities and color of the diffuse stellar population co-spatial to the point source, based on the best-fitting model (Fig.~\ref{fig:profile_fit}), with a virtual aperture of $0.5\arcsec\times0.5\arcsec$ (the same that was used for the \hst/STIS spectra extraction). In the epoch the images were taken (2020) the diffuse stellar emission accounted for less than 5\% of the total FUV emission in the innermost pixels; its color points to a mean stellar population age between 200-300 Myr, indicating some recent diffuse star formation, consistent with the relatively bright diffuse FUV emission in the central kiloparsec of the galaxy, as seen from the F140LP image in Fig.~\ref{fig:profile_fit}.

\subsection{\hst/STIS UV Spectra}\label{sec:spec}

In this section, we present an independent analysis of the near and far UV spectra of \target\ from 2014 and 2018, covering a combined wavelength range from 1150\AA\xspace to 3150\AA. Our focus is on the intrinsic shape and variability of the emission. To focus on the nuclear emission, we use a custom extraction region in calSTIS\footnote{\href{https://hst-docs.stsci.edu/stisdhb/chapter-3-stis-calibration/3-1-pipeline-processing-overview}{Link to calSTIS.}} with a 0.5x0.5\arcsec size in the along- and cross-dispersion directions respectively (see \S\ref{app:reduction} for more details). An aperture correction was used to derive the flux density.
We remove the region around observer-frame HI line wavelength (appearing at the blueshifted 1170-1210\AA wavelength range, in Fig.~\ref{fig:spec}) which is dominated by geo-coronal airglow lines\footnote{\href{https://www.stsci.edu/hst/instrumentation/cos/calibration/airglow}{HST Geo-Coronal Airglow}}.

Although the signal-to-noise (S/N) ratio is high, no strong broad emission lines typically seen in Seyfert galaxies and quasars \citep[][]{Shull2012,Stevans2014} are detected. In this sense, the UV spectrum is mostly `featureless', in comparison with AGN. This absence of lines is unsurprising, as \target\ lacks many of the classic AGN features--not only broad optical emission lines (indicating the absence of a mature broad line region, \citealt{Wevers2022}), but also no hot dust ``torus" emission nor hard X-ray corona emission \citep[][]{Miniutti2019}. A comparison to optically selected TDEs may also be warranted: their broad emission lines, usually, used to tell them apart from other optical transients, are associated with their early time ($\leq$ 1 year) `UV/optical excess' emission, whose origin is still debated but is known not to be power by direct disk emission; these lines and this emission component are not present at their later times ($>$ 1 year) when pure disk emission dominates the multi-wavelength TDE emission \citep{Charalampopoulos2022,vanvelzen19_late_time_UV, Mummery2024}. If \target does indeed have a TDE origin, then this early-time UV/optical emission phase was never observed in \target, and all the data presented in this work were taken many years after this initial phase.

The highest equivalent width (EW) features are the two emission features at 1215\AA\,(Ly$\alpha$) and at 1242/1238\AA\,(N\,V), and the (O\,I) doublet absorption feature at 1302/1305\,\AA. A detailed analysis of these features is beyond the scope of this paper, analysis of the UV emission lines in \target was only investigated by \citet{Sheng2021}. However, we caution against some of the assumptions made in \citet{Sheng2021} which should affect their main conclusions: (i) their intrinsic UV spectral shape is inconsistent with our findings (see below); and (ii) the origin of the narrow emission lines, which are \textit{assumed} as being related to debris from the potential TDE in \target; rather than the interstellar medium (ISM) gas ionized by the disk continuum, and/or by the nuclear star formation. The later interpretation, contrary to the \citet{Sheng2021} interpretation, seems to be more aligned with recent findings of a compact narrow-line region (NLR) reported by \citet{Patra2023}.

In Fig.~\ref{fig:spec}, we present the two spectra corrected for both Galactic and intrinsic extinction, using the same color excess and extinction laws as described in \S\ref{sec:image} (with a central value of $E(B-V) = 0.1$ for the host attenuation). This reveals the intrinsic shape of the UV emission from the source. The subdominant contribution from stellar emission to the overall UV spectra is also shown in green for comparison.
In Fig.~\ref{fig:spec}, we also display a $\nu F_{\nu} \propto \nu^{4/3}$ function, characteristic of the `mid-frequency range' in a standard accretion disk spectrum. It is evident that the shape of the UV spectra is not far from this analytical description, although a more detailed fitting - simultaneously accounting for the stellar contribution, is necessary. Such analysis will be presented in  \S\ref{sec:SED}.

It is useful to compare these results with those reported by \citet{Sheng2021}, where the authors found the UV spectra was best-fitted with $F_{\lambda} \propto \lambda^{-1.7}$ ($\nu F_{\nu} \propto \nu^{0.7}$), which is significantly shallower than the spectrum shown in Fig.~\ref{fig:spec}. This discrepancy is easily explained: \citet{Sheng2021} applied only Galactic extinction correction, neglecting intrinsic dust attenuation. This is an inappropriate approach given that the galaxy is known to be gas-rich and have active star formation, hence has a non-negligible amount of dust \citep[][also Fig.\ref{fig:profile_fit}]{Saxton2011,Miniutti2013,Wevers2022,Wevers2024}, with intrinsic $E(B-V)$ estimates easily obtainable from the Balmer decrement in a nuclear optical spectrum.

We now turn to the question of variability, as both \citet{Miniutti2019} and \citet{Sheng2021} reported no significant differences between the 2014 and 2018 spectra. In the middle panel of Fig.~\ref{fig:spec}, we zoom in on the FUV portion of the spectrum, where the S/N is highest. A slightly higher flux is apparent in the 2014 spectrum; however, individual bins rarely differ by more than $3\sigma$ between the two epochs.

To obtain an statistically meaningful assessment of the variability we ran a   Monte Carlo simulation of $10^4$ steps, for each of the two epochs, where in each interaction we compute the integrated FUV flux given by:

\begin{equation}\label{eq:fuv}
{\rm Flux}_{\rm FUV} = \int^{1650 \AA}_{1150 \AA} F_{\lambda}(\bar{F}, \sigma_{\bar{F}}){\rm d} \lambda
\end{equation}

\noindent Here, $F_{\lambda}(\bar{F}, \sigma_{\bar{F}})$ represents a flux density realization, randomly generated from a Gaussian distribution centered on the measured flux ($\bar{F}$) with a dispersion based on the flux uncertainty ($\sigma_{\bar{F}}$). For each realization, we recorded the total FUV flux ratio between the 2014 and 2018 observations. The resulting distribution of flux ratios is shown in the right panel of Fig.~\ref{fig:spec}. While the relative flux calibration and repeatability capabilities of \hst are excellent, they are not perfect. \citet{Bohlin2019} estimated that repeated observations of standard stars with the STIS/G140L setup typically have a systematic uncertainty of about 1\% in their relative flux ratios, with occasional outliers reaching up to 3\%. However, even the most conservative estimates of the systematic uncertainties cannot account for the $9 \pm 2$\% higher flux observed in the 2014 FUV spectrum compared to that of 2018. Instead, our simulations show that 99.9\% of the realizations yield 2014/2018 flux ratios greater than 1.03, supporting the presence of statistically significant intrinsic flux variability.

Such mild variability over four years, is not surprising, instead, this `UV plateau phase' is actually predicted by time-dependent thin disk theory 
\citep{Cannizzo1990,Mummery2020} and is now well-established in the observational behavior of the broader TDE population \citep{vanvelzen19_late_time_UV,Mummery2024}. This plateau is a result of the simultaneous canceling of two competing effects, cooling  (which would reduce the flux) and expansion (which would increase the flux) of the compact accretion disk formed in the aftermath of the TDE. This behavior will be explored further in the next sections.

\begin{figure*}[t!]
	\centering
	\includegraphics[width=0.95\textwidth]{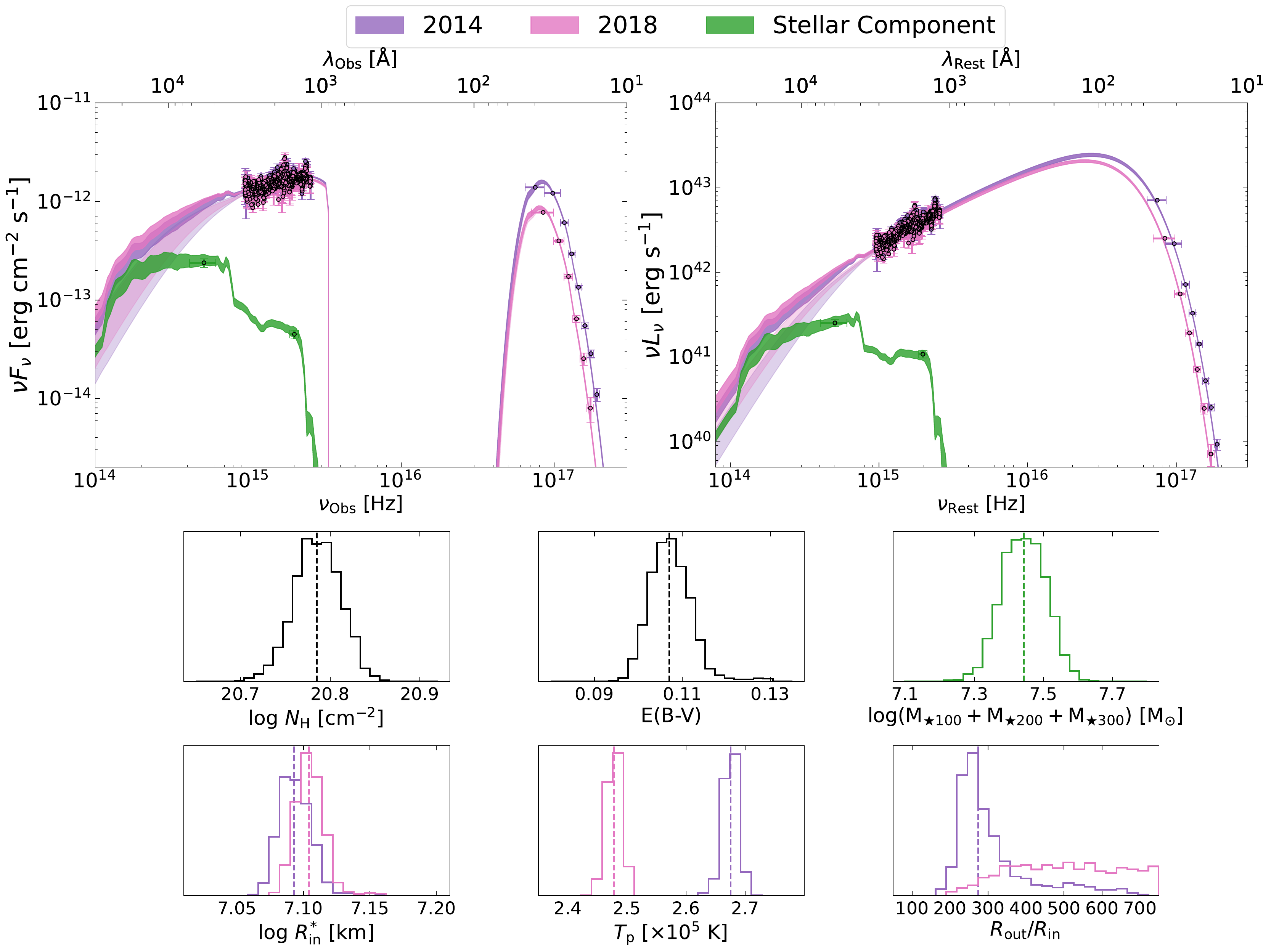}

	\caption{
    Results of the nested sampling fit of broad-band data, using \texttt{diskSED}. Purple colors refer to 2014, pink to 2018, and green to the stellar component. For the two epochs the darker contours are total models (disk+stars), while the lighter contours are the disk-only emission. {\bf Top Left:} observed flux models (without any extinction/absorption correction) overlaid on the observed UV/optical data and the unfolded X-ray spectra. {\bf Top Right:} intrinsic luminosities (all absorption/extinction corrections), with the data points unfolded to the median values of the parameter posteriors. {\bf Bottom}: Marginalized posteriors of the free parameters. Vertical lines show the median values of the posterior distributions that have fully converged.}
    \label{fig:SED}
\end{figure*}

\section{Fitting of contemporaneous X-ray and UV spectra}\label{sec:SED}

Given that the nuclear UV point source cannot be of stellar origin, it is natural to expect it to be related to the quiescent X-ray emission. The soft/thermal X-ray emission in \target and in TDEs in general, is usually well described by the inner emission of a standard radiatively efficient thin disk. If that is indeed the case then the UV emission should be associated with the mid-to-outer parts of the disk. The flux and spectral shape of the UV/optical SED can (when fitted simultaneously to the X-ray data) therefore, constrain the extension of the disk 
\citep{Mummery2020,mummery2024fitted,Nicholl2024,Guolo_Mummery2024}. 

\subsection{Fitting Setup and Modeling in the Newtonian Limit}\label{sec:diskSED}

Given the TDE-like nature of \target's emission, with original disruption likely just a few years before our first epoch \citep{Miniutti2023_longterm}; the accretion disk should be somewhat compact (see further discussion in \S\ref{sec:discus}). Standard spectral models available in \texttt{XSPEC}, developed for X-ray binaries and/or AGN, are inadequate for such fitting as they assume a very extended (or formally infinite) disk size.

Recently, models that include the outer edge/radius of the disk (\Rout) as a free parameter, alongside standard accretion disk parameters, have been incorporated into the \texttt{pyXspec} X-ray fitting package by \citet{Guolo_Mummery2024}. These models are specifically designed for broad-band analysis (combining X-ray spectra with UV/optical spectra and/or photometry), allowing both the inner and outer disk properties to be marginalized during fitting.
Our first goal is to investigate whether the broad-band SED can be self-consistently described by a standard disk model. To this end, we begin our analysis with the simplest model, \texttt{diskSED}, as presented in \citet{Guolo_Mummery2024}. 
The \texttt{diskSED} model implements a classical thin disk framework \citep{Shakura1973} with a null-stress boundary condition at the inner edge. It incorporates radiative transfer effects in the disk atmosphere, accounting for electron scattering and metal opacity through a temperature-dependent color correction factor \citep[$f_c$,][]{Shimura1995,Hubeny2001}. In this model, photon propagation is treated in the Newtonian approximation, assuming straight-line trajectories from the emission region to the observer's image plane. Consequently, disk inclination ($i$) and black hole spin ($a$) are not free parameters and cannot be marginalized in the fitting process.
The model has three free parameters: the peak physical temperature of the disk ($T_{\rm p}$), \Rins\ ($\equiv R_{\rm in}\sqrt{\cos i}$), where $R_{\rm in}$ is the disk's inner radius and $i$ its inclination relative to the observer, and the dimensionless size of the disk (\Rratio).

To model dust attenuation intrinsic to the host galaxy, we use the \texttt{reddenSF} model \citep{Guolo_Mummery2024}, which employs the \cite{Calzetti2000} attenuation law from 2.20 $\mu m$ to 0.15 $\mu m$ and its extension down to 0.09 $\mu m$ as described in \citet{Reddy2016}. The free parameter of the \texttt{reddenSF} model is the color excess E(B--V).
The Galactic extinction is modeled with \texttt{XSPEC}'s native \texttt{redden}, which employs \citet{Cardelli1989}'s law.
Modeling of the neutral gas absorption is carried with two instances of the \texttt{phabs} model, one Galactic and another redshifted to the host frame.
The intrinsic part of the model is shifted to the source rest frame using \texttt{zashift} with $z = 0.018$. 

The UV spectra of \target have some minor, but non-negligible,  contribution from the underlying stellar population, for which we have fluxes/color in two bands (see green points in Fig.~\ref{fig:point_source} and \ref{fig:spec}). Simultaneous modeling of this contribution is necessary. We again use the SSP models from \citet{Maraston05}, we integrated these SSP's SED into the fitting as a model (which we will call \texttt{stellarPop}). Based on the color of the stellar population component (Fig.~\ref{fig:point_source}), we allow for SSPs of age 100, 200 and 300 Myr to be used in the fitting. The stellar masses of the each of the components are the free parameters ($M_{\bigstar,{\rm age/Myr}}$).

The SED modeling, fitting, and full parameter space sampling is performed using the Bayesian X-ray Analysis software (BXA) version 4.0.7 \citep{Buchner2014}, which integrates the nested sampling algorithm \texttt{UltraNest} \citep{Buchner2019} with the fitting environment \texttt{PyXspec} - which is the \textit{Python} version of the classical \texttt{XSPEC} \citep{Arnaud_96}.  UV/optical photometry of the stellar population and the UV spectra are added to \texttt{PyXspec} (without extinction correction) using the ``ftflx2xsp" tool available in HEASoft v6.33.2 \citep{Heasarc2014}, which generates the response file for the fitting package. To be conservative on our conclusions that rely on the UV spectra fluxes, we added a 3\% systematic uncertainty to the UV spectra when loading into the fitting package. While X-ray spectra can be fit using Poisson statistics in their native instrumental binning, \texttt{XSPEC} does not support fitting UV/optical data with Poisson statistics. Therefore, the X-ray spectra are binned using an `optimal binning' scheme \citep{Kaastra2016}, also ensuring that each bin contains at least 10 counts (i.e., \texttt{grouptype=optmin} and \texttt{groupscale=10} in \texttt{ftgrouppha}), and the simultaneous X-ray plus UV fit is performed using Gaussian statistics.

In summary, we have three data sets to be simultaneously fitted: two epochs of X-ray plus UV spectra, and the stellar (diffuse) component's photometry in two filters. In \texttt{XSPEC} notation, the two epochs of X-ray and UV spectra are therefore fitted with \texttt{phabs$\times$redden$\times$zashift(phabs$\times$reddenSF$\times$
(stellarPop + diskSED))}, while the stellar component's photometry is fitted with \texttt{redden$\times$zashift
(reddenSF$\times$stellarPop)}.

The Galactic X-ray neutral gas absorption is fixed to the Galactic hydrogen equivalent column density equals to $N_{H, G} = 2.3 \times 10^{20}$ cm$^{-2}$ \citep{HI4PI2016} and the Galactic extinction is fixed to E(B-V)$_{G}$ above. The three parameters of \texttt{diskSED} (\Rins, \Tp, and \Rratio) are free to vary independently in each two epochs. The other parameters (E(B--V), $N_{\rm H}$, and $M_{\bigstar,{\rm 100}}$, $M_{\bigstar,{\rm 200}}$, $M_{\bigstar,{\rm 300}}$) are free to vary but are tied between the three data sets, as they are intrinsic to the host galaxy and should not vary with time. Uniform (of log-uniform) priors are assumed for all the free parameters.

The results of the nested sampling fit are shown in Fig.~\ref{fig:SED}. The bottom panel shows the 1D projection of the posterior distributions. The full posteriors of all parameters are shown in Appendix \S\ref{app:post}. In the left upper panel of Fig.~\ref{fig:SED}, we show the observed flux models (without extinction/absorption corrections) overlaid on the observed UV photometry and the unfolded X-ray spectra. The right panel shows the intrinsic luminosities (with both Galactic and intrinsic absorption/attenuation corrections), with the data points unfolded to the median values of the parameter posteriors. In the upper right and left panels, green is the stellar component, while 2014 and 2018 SED are respectively purple and pink.

Given the limited observational constraints on the stellar component — specifically, just two photometric bands and its minor contribution to the UV spectra — strongly constraining its properties is challenging. Distinct combinations of the three available SSPs appear capable of reproducing the stellar emission in the inner 0.5\arcsec$\times$0.5\arcsec of \target's host (see Appendix \ref{app:post}).  However, a total stellar mass of log($M_{\bigstar}$) $\approx 7.4\pm 0.1$ \msun is well-constrained (green histogram in Fig. \ref{fig:SED}). The selection of only these three ages does not imply the absence of older stars; rather, it reflects that, with the available photometry, only the contributions of stars young enough to emit at optical/blue and UV wavelengths can be constrained, as older stars ($\geq$ 1 Gyr) contribute negligible light at wavelengths bluer than the red/optical and near-infrared bands. The intrinsic color-excess, $E(B-V) = 0.11 \pm 0.01$, recovered from the full SED fitting, is consistent with the completely independent estimates obtained by \citet{Wevers2024} from the Balmer decrement of \target's optical nuclear spectra, as mentioned before.

The \Rins\ parameter exhibited no significant evolution between the two epochs, supporting our earlier suggestion that both the X-ray and the majority of the UV emission originate from the disk. While observables such as UV and X-ray fluxes have evolved, they have done so in a manner consistent with the disk parameters expected to remain constant over time (\Rin\ and $i$) remaining unchanged. This consistency indicates that an evolving accretion disk provides a robust description of the SED.

Among the disk parameters, the inner disk temperature (\Tp) shows the most significant evolution from epoch two epochs. The conditional posterior probability for  cooling, i.e., $P(T_{\rm p}^{2014} > T_{\rm p}^{2018} | \rm{data}) = 1.0$ indicating that the cooling of the disk, by $\sim$10\%, is recovered at extremely high significance. This cooling is a fundamental prediction of time-dependent disk evolution theory \citep[e.g.,][]{Cannizzo1990, Mummery2020}. While this cooling had already been confirmed through analyses of X-ray spectra alone \citep{Miniutti2023_longterm} it is reassuring to observe this evolution when simultaneously fitting the X-ray and the UV.

Finally, we examine the disk size results. The 2014 SED parameter \Rratio has well-converged to a value of \Rratio $\approx 280_{-40}^{+100}$. This means that even in the simplest Newtonian limit, we can constraint a finite somewhat compact disk for \target, in the first epoch. However, for 2018, the sampling in the fit could not constrain the upper limit of \Rratio, with the posterior distribution remaining essentially flat for \Rratio $\gtrsim 330$. Thus, only a lower limit of \Rratio $\gtrsim 300$ (90\% posterior) could be determined. The reason for the differing constraint power, despite the same wavelength coverage, is that as the disk cools, its SED shifts to lower frequencies, moving the characteristic break between the mid-frequency range and the Rayleigh-Jeans tail of the SED to frequencies for which we lack contemporaneous data coverage (see Figure 2 in \citet{Guolo_Mummery2024} for illustration), which weakens the convergency capabilities of the sampling.

More physically meaningful than \Rratio\ and \Rins\ would be estimates of $R_{\rm out}$ (in units of the gravitational radius, $R_{\rm g}$) and \Rin. The former would allow for direct comparison with $R_{\rm orb}$, while the latter, when associated with the innermost stable circular orbit (ISCO), is directly correlated with the black hole mass (\Mbh), and would allow for \Mbh estimates. However, given that in the Newtonian limit of \texttt{diskSED} $i$ and $a$ cannot be marginalized, $R_{\rm out}/R_{\rm g}$ and \Mbh estimates can only be obtained by assuming \textit{ad hoc} values (or probability distributions) of $a$ and $i$ \citep[see e.g.,][]{Wevers2025,Guolo_Mummery2024}.
Instead, given the high-quality available data on \target, i.e., two epochs of very high S/N X-ray spectrum and wide UV coverage, estimates on inclination and spin may be obtainable via introduction of relativistic effects into the disk modeling. This is the focus of the next section.

\subsection{Relativistic Disk Modeling}\label{sec:kerrSED}

In a more realistic Universe, photons traveling through Kerr spacetime around a black hole undergo both energy shifts (as measured by a local observer) and trajectory deviations from straight lines. Phenomena such as gravitational redshifting, gravitational lensing, and Doppler boosting naturally arise, depending on the spin of the black hole and inclination of the disk relative to the observer. However, except in special viewing geometries, these relativistic effects cannot be computed analytically and require numerical ray tracing.
The \texttt{kerrSED} model implementation \citep{Guolo_Mummery2024} extends the classical thin disk framework by including relativistic effects. Similar to \texttt{diskSED}, \texttt{kerrSED} employs a color-corrected standard thin disk model with a null-stress inner boundary condition but it incorporates full relativistic ray tracing ``on the fly" using the numerical algorithm presented in \citet{mummery2024fitted}. This approach breaks the perfect degeneracy between $i$, $a$, and other disk parameters present in the Newtonian case, enabling constraints on inclination and spin provided high-quality data is available.

The \texttt{kerrSED} model has five free parameters: \Rin, \Tp, \Rratio, $a$ and $i$. We repeat the fitting approach described in \S\ref{sec:diskSED}, simultaneously fitting both epochs of X-ray and UV spectra along with the stellar population component. Parameter constraints are kept consistent with the previous section, with the exception that \Rin\ is now tied between the two epochs, as we have already demonstrated that the SED can be described by a disk model. Additionally, the two new parameters, $a$ and $i$, are also tied between epochs\footnote{The disk inclination angle, $i$, could potentially vary slowly over time if the disk is precessing \citep{Miniutti2024}. However, with only two epochs of full SED, such effects cannot be incorporated into our modeling. Consequently, our inferred $i$ values should be interpreted as a ``time-average and S/N-weighted inclination" over the period of the two observations.}. All other aspects of the nested sampling, including the use of uniform or log-uniform priors, remain unchanged.

\begin{deluxetable}{l|c|c}
\label{tab:par}
\tabletypesize{\small}
\tablecaption{Inferred parameters for \texttt{kerrSED} fitting of \target.}
\tablehead{ \colhead{\rm Parameter} &  \colhead{High Spin Mode} &  \colhead{Low Spin Mode} \\ 
 \colhead{} &  \colhead{($a \geq 0.75$)} &  \colhead{($a < 0.75$)} }
\startdata
$a$                                     & $0.95^{+0.02}_{-0.05}$               & $-0.2^{+0.5}_{-0.4}$         \\
$i \, [{\rm deg}]$                      & $59^{+4}_{-2}$                       & $41^{+7}_{-6}$                  \\
$R_{\rm in} \, [{\rm km}]$              & $1.5^{+0.1}_{-0.1} \times 10^7$      & $1.2^{+0.1}_{-0.1} \times 10^7$ \\
$M_{\rm BH} \, [M_{\odot}]$             & $7.5^{+1.0}_{-1.5} \times 10^6$      & $1.5^{+0.5}_{-0.3} \times 10^6$ \\
$T_{\rm p}^{\rm 2014} \, [{\rm K}]$     & $2.72^{+0.05}_{-0.03} \times 10^5$   & $2.77^{+0.07}_{-0.06} \times 10^5$ \\
$T_{\rm p}^{\rm 2018} / T_{\rm p}^{\rm 2014}$ & $0.91 \pm 0.01$               & $0.93 \pm 0.01$               \\
$R_{\rm out}^{\rm 2014} \, [R_{\rm g}]$ & $1200^{+400}_{-250}$                 & $4100^{+1550}_{-1280}$          \\
$R_{\rm out}^{\rm 2018} / R_{\rm out}^{\rm 2014}$ & $1.15 \pm 0.02$             & $1.11 \pm 0.01$                 \\
$L_{\rm Bol}^{\rm 2014}$ [erg s$^{-1}$]          & $5.5_{-0.2}^{+0.5} \times 10^{43}   $           & $3.8_{-0.3}^{+0.3} \times 10^{43} $          \\
$L_{\rm Bol}^{\rm 2018}$ / $L_{\rm Bol}^{\rm 2014}$ & $0.69 \pm 0.01$  & $0.74 \pm 0.01$                 \\
$\lambda_{\rm Edd}^{\rm 2014}$          & $0.05^{+0.01}_{-0.01}$               & $0.20^{+0.05}_{-0.05}$          \\
$\lambda_{\rm Edd}^{\rm 2018} / \lambda_{\rm Edd}^{\rm 2014}$ & $0.69 \pm 0.01$  & $0.74 \pm 0.01$             
\enddata
\tablecomments{Central values represent the medians of the respective mode of the posteriors, with uncertainties corresponding to their 68\% credible intervals.}
\end{deluxetable}

\begin{figure}[h!]
	\centering
	\includegraphics[width=\columnwidth]{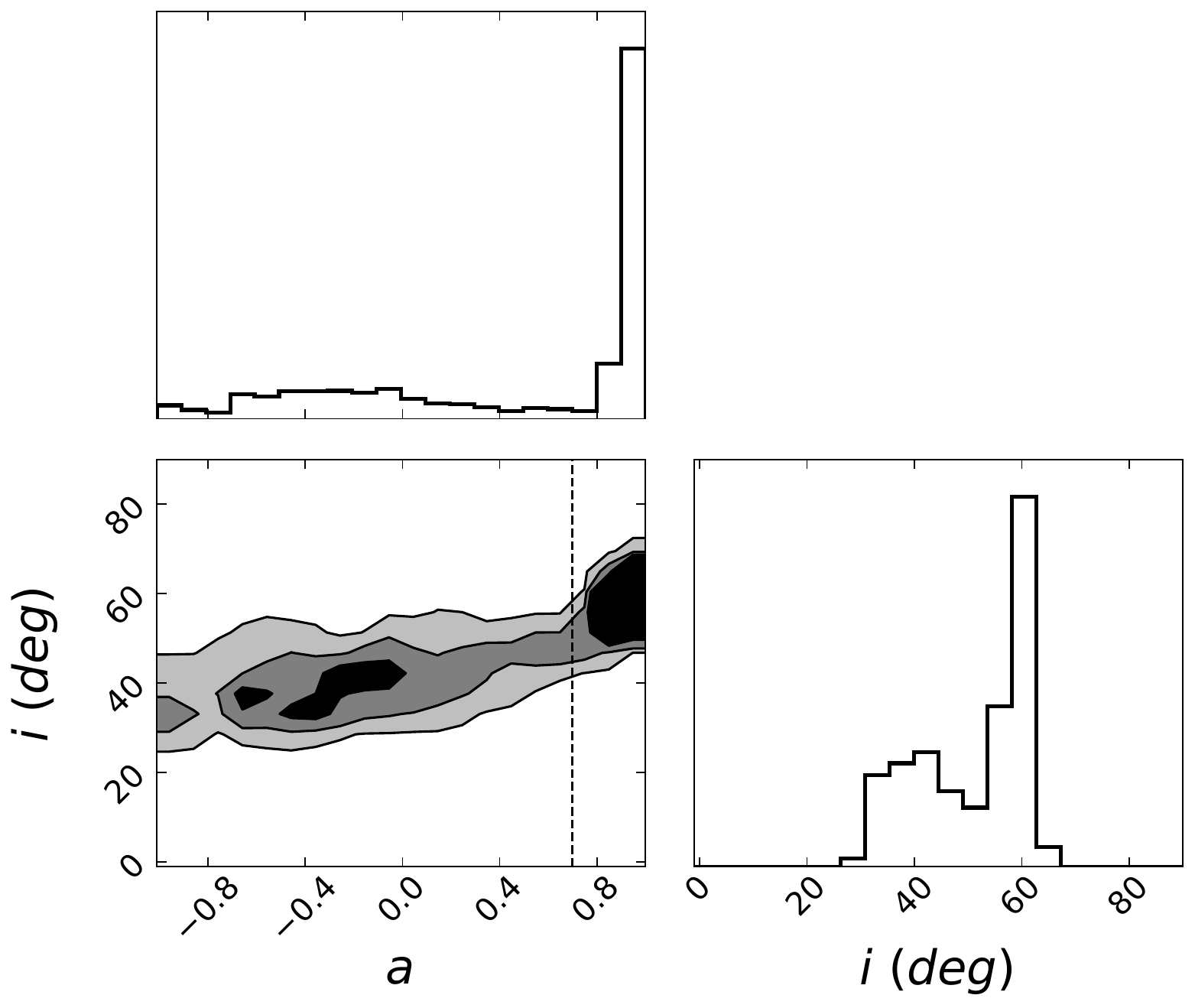}
 
	\caption{Corner plot showing the bimodal posterior of the spin ($a$) and inclination ($i$), for the \texttt{kerrSED} fitting of \target. Contours are 68\%, 90\% and 99\% of the marginalized posterior. Dashed line shows the criteria used to separate the two modes.}
    \label{fig:a_x_i}
\end{figure}

The SED plots, analogous to Fig.~\ref{fig:SED}, and the complete parameter posteriors from the relativistic fitting are presented in Appendix \S\ref{app:post}. Most of the recovered parameters (e.g.,$M_{\star}$, E(B-V), $N_{\rm H}$), are consistent with those recovered in the Newtonian limit. First, we shall focus on the constraints obtained for the two previously unknown parameters, $a$ and $i$. Their marginalized posteriors are shown in Fig.~\ref{fig:a_x_i}. As the figure illustrates, the relativistic model effectively constrains the inclination of \target's disk: 99\% of the $i$ probability distribution is contained within $31^\circ \lesssim i \lesssim 63^\circ$.

\begin{figure*}[]
	\centering
	\includegraphics[width=0.9\textwidth]{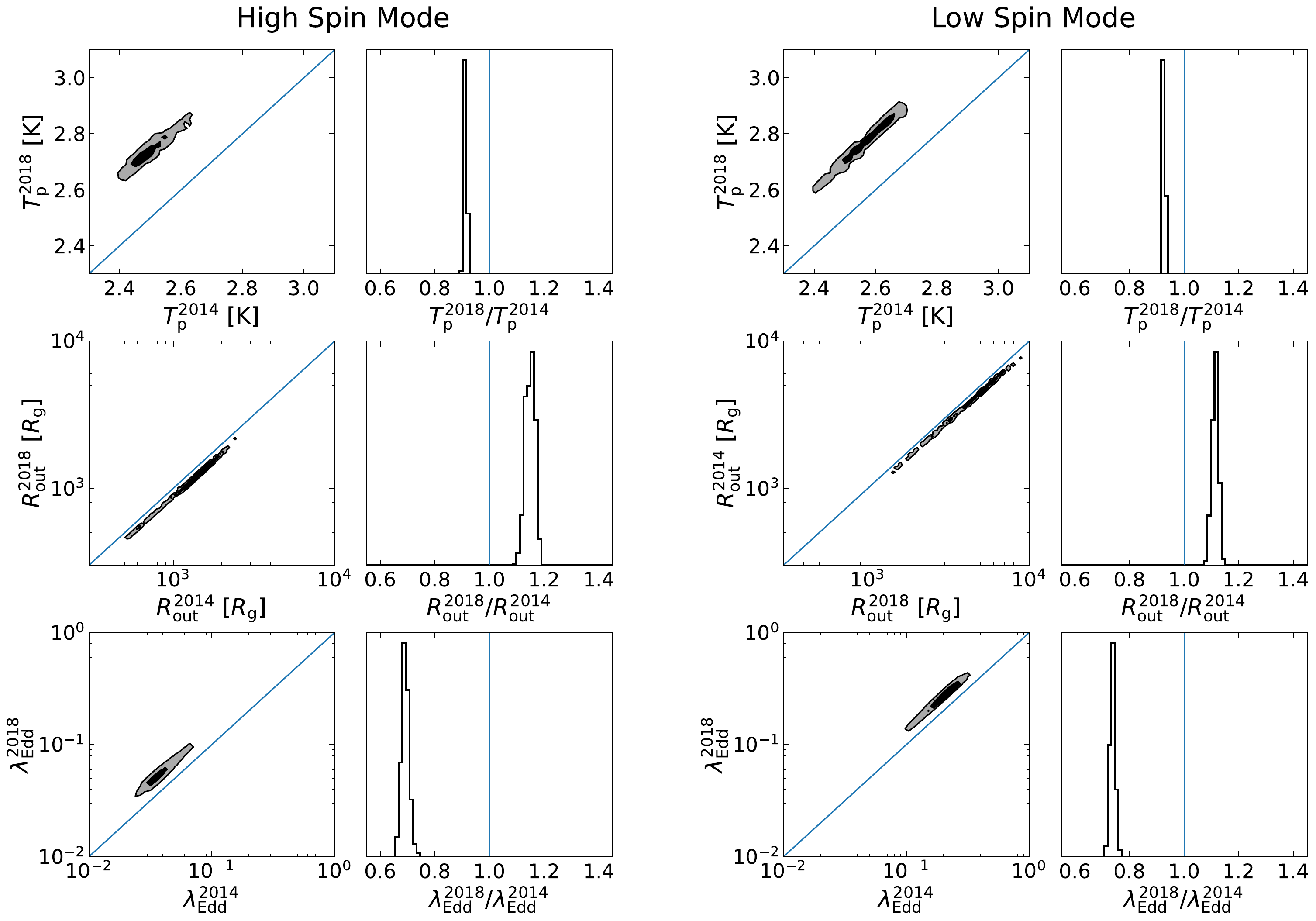}
 
	\caption{Posterior distributions of the time-evolving properties of \target's disk, as fitted with \texttt{kerrSED}, are shown for the two solution modes. The top, middle, and bottom panels respectively display the physical peak disk temperature ($T_{\rm p}$), the outer disk radius ($R_{\rm out}$) in units of $R_{\rm g}$, and the Eddington ratio ($\lambda_{\rm Edd}$). The left panels illustrate the 2D posteriors for each parameter across the two epochs (with contours representing 68\% and 99\% of the posterior), while the right panels present the probability distribution functions for the parameter ratios between the 2018 and 2014 epochs. The results clearly demonstrate simultaneous cooling and expansion of the disk from 2014 to 2018.}
    \label{fig:cool_expansion}
\end{figure*}
In contrast, no range of the spin parameter ($a$) can be excluded with high confidence, although the probability is primarily concentrated in two regions: high prograde spins and another with null to slightly retrograde spins. This bimodal posterior distribution is not limited to $a$ but is also observed in $i$ and other parameters, such as \Rin\ and \Rratio\ (see Appendix \S\ref{app:post}). Consequently, secondary parameters of interest, such as \Mbh\ and $R_{\rm out}/R_{\rm g}$, also exhibit bimodal distributions. This indicates that two distinct sets of solutions can maximize the likelihood similarly well, for the available data. This bimodal posterior can be improve, and perhaps became mono-modal, by e.g., including more epochs of X-ray spectra and/or performing a fully time-depending fitting of the flow evolution; but this will be the subject for future work.

While the data do not allow precise determination of the spin, the relativistic model still recovers significant new physical constraints. Notably, the $a \times i$ parameter range consistent with the data is much smaller than the prior. For instance, the outermost gray contour in Fig.~\ref{fig:a_x_i}, marking the 99\% credible interval of the posterior, encompasses less than 30\% of the prior area.

We follow the approach of \citet{Vieira2023}, to analyze bimodal posterior distributions by separating the solutions using an (arbitrary) cut in the main parameter driving the bimodality — in our case, the spin. Specifically, we divided the full posterior into two subsets: interactions of the sampler resulting in $a \geq 0.75$ (the high spin mode) and those with $a < 0.75$ (the low spin mode). Now that both \Rin and $a$ were marginalized and distribution for both are available, we can obtain corresponding probability distributions for the black hole mass, by simple identifying \Rin with the ISCO, such that

\begin{equation}\label{eq:M_BH_kerr}
    M_{\rm BH} = \frac{ R_{\rm in} c^2  }{\gamma(a) G}.
\end{equation}

\noindent where $G$ is the gravitational constant and $\gamma(a)$ is the ISCO location in units of $R_{\rm g}$ \citep{Bardeen72}, which correspond to $\gamma(-1) = 9$ , $\gamma(0) = 6$ and $\gamma(1) = 1$.

Following this approach, we obtain \Mbh $ = 5.5^{+1.1}_{-1.5} \times 10^6$ \msun (high spin mode) and \Mbh $= 1.5^{+0.5}_{-0.3} \times 10^6$ \msun (low spin mode). These values are consistent with host-galaxy scaling relations. The nuclear stellar velocity dispersion of \target's host is $\sigma_* = 63 \pm 4$ km s$^{-1}$ \citep{Wevers2022}, which translates into \Mbh\ estimates of ${\rm few} \times 10^{5} - {\rm few} \times 10^{6}$ \msun (for the central value only, not adding the intrinsic dispersions)\ using distinct \Mbh$-\sigma_*$ relations. These relations, however, have systematic $1\sigma$ dispersions of approximately ${\pm\, 0.5 \ \rm dex}$. Thus, even considering the full spin posterior distribution, our \Mbh\ estimates are more precise than those derived from host-scaling relations, as our 99\% credible interval is (1, 9)$\times 10^{6}$ \msun, more importantly, our solutions for distinct black hole masses have distinct associated spins and inclination distributions (Table~\ref{tab:par}).

Similarly, we derive the probability distribution for \Rout/$R_{\rm g}$ simply given by $\left( \frac{R_{\rm out}}{R_{\rm in}} \right)\times\gamma(a) \). For 2014, we find \( R_{\rm out}^{\rm 2014} = 1208^{+424}_{-250} \, R_{\rm g} \) (high spin mode) and \( R_{\rm out}^{\rm 2014} = 4096^{+1548}_{-1280} \, R_{\rm g} \) (low spin mode). For the 2018 epoch, we find that the disk is a factor of $R_{\rm out}^{\rm 2018}/R_{\rm out}^{\rm 2014} = 1.15 \pm 0.02$ larger, for the high spin mode; or similarly a factor of $R_{\rm out}^{\rm 2018}/R_{\rm out}^{\rm 2014} = 1.11 \pm 0.01$ larger for the low spin mode.

\begin{figure*}[t]
    \centering
    
    \includegraphics[width=0.9\textwidth]{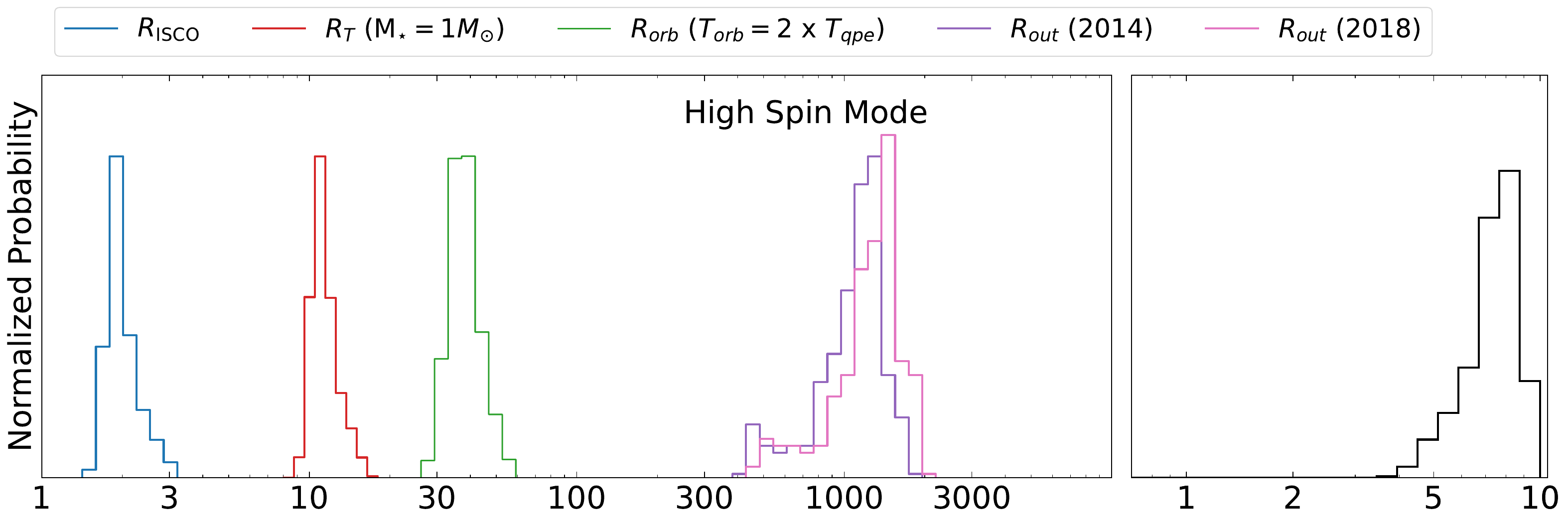}\\
    \includegraphics[width=0.9\textwidth]{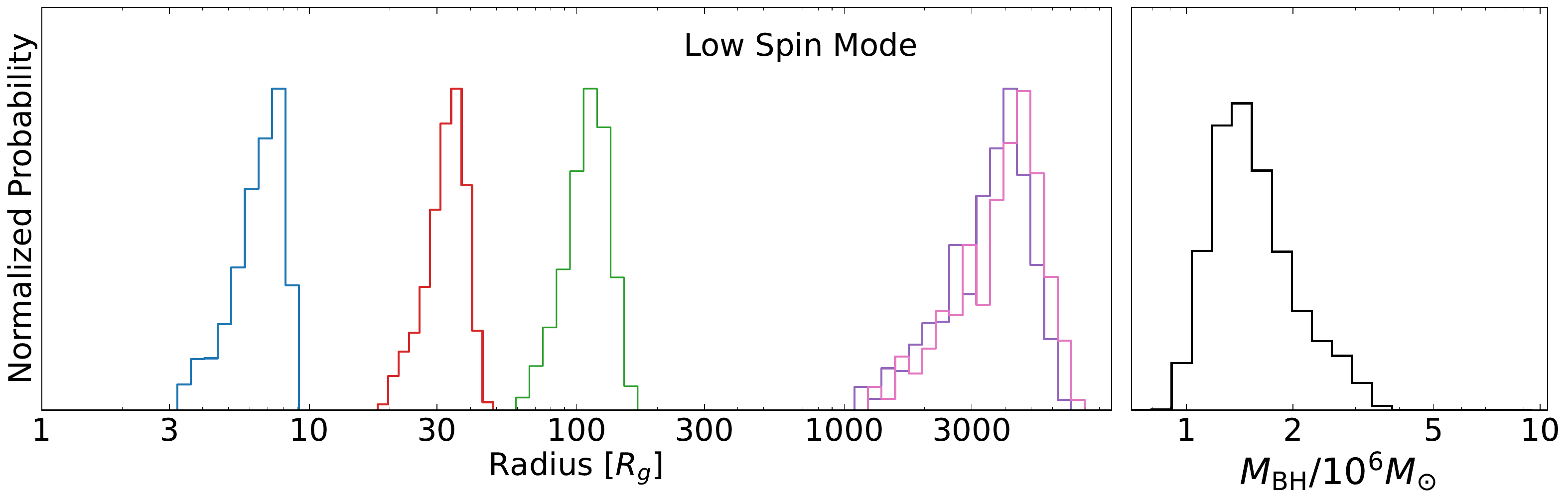}
    \caption{Probability distribution functions (PDFs) for the inferred parameters of \target\ using the relativistic \texttt{kerrSED} disk model. Due to the bimodal nature of the final posterior distribution (driven by the not fully constrained spin parameter, $a$), the solutions are divided into two modes (see Table \ref{tab:par}). The top panels display the PDFs for the `high spin mode', while the bottom panels show those for the `low spin mode'. The right panels present the black hole mass (\Mbh), while the left panels illustrate the radial distributions of key parameters. The orbiter radius ($R_{\rm orb}$, green), relevant for disk/orbiter interaction QPE models, is calculated assuming a circular orbit and $T_{\rm QPE} = 9$ hr. For reference, the tidal radius ($R_{\rm T}$, red) is shown, based on the \Mbh\ distribution and assuming a 1\msun main-sequence star.}
    \label{fig:radii_profile} 
\end{figure*}

A summary of all parameters and uncertainties for the two solution modes is presented in Table~\ref{tab:par}. \Mbh and $R_{\rm out}/R_{\rm g}$ probability distributions are shown in Fig.~\ref{fig:radii_profile}. While the probability distributions and the 2D posteriors of the time evolving properties are shown in Fig.~\ref{fig:cool_expansion}. 

Using the Bayesian framework applied here and the marginalized posteriors for both \Tp\ and \Rout\ at the two epochs, we can ask what is the probability that the disk has cooled and what is the probability that the disk has expanded. Formally, we find that \( P(T_{\rm p}^{2014} > T_{\rm p}^{2018} \mid \mathrm{data}) = 1.0 \) and \( P(R_{\rm out}^{2018} > R_{\rm out}^{2014} \mid \mathrm{data}) = 1.0 \). This can be easily visualized by top and middle panels of Fig.~\ref{fig:cool_expansion}. This indicates that, in every sampling iteration, the solutions with both disk cooling and expansion are the solutions that maximize the likelihood. Given our flat prior distributions (\(\pi(T_{\rm p}^{2014} > T_{\rm p}^{2018}) = 0.5\) and \(\pi(R_{\rm out}^{2018} > R_{\rm out}^{2014}) = 0.5\)), the overwhelming evidence for cooling/expansion therefore is reflecting an inherent property of the data, when such evolution is allowed into the modeling, i.e., when both inner disk temperature and outer radius are free parameters of the model. We will further discuss these findings in the context of \target's nature in next section.

With the multi-wavelength fitting performed, we can now obtain more precise estimates of the bolometric disk luminosity ($L_{\rm Bol}$) compared to estimates based solely on X-ray data. For the two epochs, we find $L_{\rm Bol}^{2014} = \rm{few} \times 10^{43}$ \ergs, with slightly deference between the modes, given the anisotropy of the relativistic disk, owing the angle-dependent relativistic effects, a decrease of approximately $30\%$ over the four-year period is observed in both modes. 
The Eddington ratios ($\lambda_{\rm Edd} = L_{\rm Bol} / L_{\rm Edd}$) however, differ more significantly mainly due to a factor of $\sim5$ difference in the median recovered $M_{\rm BH}$ values for the two modes. The corresponding values are listed in Table~\ref{tab:par} and illustrated in Fig.~\ref{fig:cool_expansion}.

\section{Discussion}\label{sec:discus}
Having recovered the system's properties through the multi-wavelength analysis, we can now explore how the parameter values and their evolution over time to constrain different models. This includes not only models for the origin of QPEs in the system but also the nature of \target's emission prior to the onset of the eruptions.

\subsection{The Nature of GSN\,069 Emission}\label{sec:gsn_tde}
One of the key differences between TDEs and standard AGN lies in the radial extent of their respective accretion flows. The characteristic size of the outer disk edge in both systems is set by the origin of the  material which is being accreted. In AGN, the accretion flow is sustained by large-scale (circumnuclear) gas transport within the host galaxy \citep[e.g.,][]{Storchi-Bergmann2019,AnglesAlcazar2021,Hopkins2024a, Hopkins2024}, and is long-lived, with the accretion flow extending to scales of at least $\mathcal{O} (10^5 \, R_g)$. 

This large-scale feeding mechanism also gives rise to the additional structures observed in AGN systems (in which the material typically exists in distinct physical states or compositions compared to the optically thick, ionized accretion flow). For example the Broad Line Region (BLR) and the dusty `torus' \citep[e.g.,][]{Peterson2004,Cackett2021,Hopkins2024c}, typically extend to even larger scales than the disk itself. 
Our inferred \(R_{\rm out}\) confidently rules out the presence of such large-scale accretion flows in \target. Combined with the absence of all other ubiquitous AGN features, as previously discussed, our findings definitively exclude earlier suggestions that \target could be a `true Type 2' AGN \citep[e.g.,][]{Saxton2011, Miniutti2013, Shu2018}\footnote{In the AGN unification model, all AGN are intrinsically similar, differing only in luminosity and producing the same broad emission line spectrum. The absence of broad emission lines in Type 2 AGN is explained by the viewing angle relative to the toroidal obscuration surrounding the central engine \citep{Antonucci1993}. In contrast, `true Type 2' AGN are hypothesized to be long-lived systems that inherently lack a broad line region (BLR). However, the existence of such AGN remains highly debated \citep{Antonucci2012}, as many initial candidates were later shown either to possess a BLR \citep[e.g.,][]{Bianchi2019} or to be TDEs.}. The position of \target's nuclear optical spectra emission line ratios in the region above the \citet{Kewley2001} line of the BPT diagram (a.k.a the `AGN region') can be explained by the formation of a compact narrow line region produce by the $\gtrsim 10$ year of high ionizing luminosity produced by \target, as shown by e.g., \citet{Patra2023} and \citet{Mummery2025}, and does not require a long-lived AGN flow.

In a TDE, however, the source of the accreting material is initially deposited very close to the MBH. Specifically, the initial disk size is set (to leading order) by the pericenter radius of the orbit of the star which was disrupted \citep{Rees1988,Cannizzo1990,Gezari2021}. For a star to be fully disrupted, this is naturally significantly closer to the black hole than the scale of circumnuclear gas feeding in an AGN. 
This tidal radius can be written as a function of the black hole and disrupted star properties: $R_T \approx R_* \left( M_{\rm BH}/M_* \right)^{1/3}$, with \(R_*\) and \(M_*\) being respectively the stellar radius and mass. For typical black hole and stellar parameters, the initial disk size should extend to a few tens to a few hundreds of \(R_g\). For example, a solar type star disrupted by a $10^6M_\sun$ MBH would have a tidal radius of $\sim 50R_g$. Clearly, TDE disks are expected to be much more compact (initially) than an AGN flow. 

This compact disk will then evolve \citep[e.g.,][]{Cannizzo1990,Mummery2020}, with material being lost through the inner edge of the flow (i.e., the disk is accreted). The accretion process conserves angular momentum, and as the total mass of the disk drops, some material must therefore be sent to larger radii to conserve the total angular momentum of the flow $(J_{\rm disk} \propto M_{\rm disk}(t) \sqrt{R_{\rm out}(t)} = {\rm constant})$. This means that while initially very compact ($R_{\rm out} \sim 10$'s $R_g$), as time elapses the disk will expand ($R_{\rm out} \sim 100$-$1000$'s $R_g$). Classical disk solutions predict a power-law time-dependence of the disk outer edge ($R_{\rm out} \propto (t/t_{\rm visc})^{3/8}$ for the canonical \citealt{Cannizzo1990} model, where $t_{\rm visc}$ is the so-called ``viscous'' timescale of the flow). As can be seen from this expression, the expansion rate of the disk is strongest at early times $v_{\rm out} \equiv  {\rm d}R_{\rm out}/{\rm d}t \sim t^{-5/8}$, meaning the initial expansion is pronounced, followed by a gradual slowing of the expansion rate (physically this is simply because the mass of the disk drops most quickly at early times, so the expansion must be rapid to conserve $J_{\rm disk}$).

An actual inference of the numerical value of \(t_{\rm visc}\) for \target's disk lies beyond the scope of this work, as it requires fully time-dependent modeling of the flow \citep[e.g.,][]{Mummery2020, mummery2024fitted}, but  will be the subject of future investigations. Nevertheless, since the time of disruption is not well known for \target our first \Rout estimates date to at least 4 years after the initial disk formation. As the expansion of the disk is most rapid in these first 4 years, an inferred \Rout of $\mathcal{O} (10^3 \, R_g)$ is certainly a plausible expectation for a TDE disk, at such late times. For example, this is just a factor of ${\rm few}$ larger than the estimated for AT2019qiz $\sim$ 4 years after disruption \citep{Nicholl2024}. 

Even without imposing a any time evolution in the disk modeling, the cooling and expansion of the flow are the preferred evolution of the disk parameters (\Tp\ and \Rout) over the four-year period analyzed here, as discussed in \S\ref{sec:SED}. This further supports the TDE nature of \target's disk, as the cooling means that the mass of the disk is decreasing with time (i.e., the disk is not in steady-state), and the expansion means the source of initial mass was within the first measured outer edge. This is further supported by the previously reported cooling of the flow, from its initial discovery through to the latest epoch analyzed here, as noted in X-ray-focused studies \citep[][]{Miniutti2019,Miniutti2023_longterm,Miniutti2023_noqpe}. 

\subsection{Implications for QPE models}
We discuss the implications of our findings for the models proposed to explain QPEs. Our aim is not to exhaustively examine every sub-variety of these models, given that the number of proposed models exceeds the number of known QPE sources. Instead, we focus on classes of models gaining traction in the literature as more data-driven properties of QPEs become better understood. 

For instance, models invoking binary MBH lensing appear to be disfavored due to the observed spectral evolution during eruptions \citep{Miniutti2023_noqpe, Arcodia2022, Arcodia2024, Giustini2024}. Similarly, models that do not involve an accretion disk at all, are also disfavored, as a radiatively efficient disk is required to account for the quiescent (in between the eruptions) emission, now with its optical/UV counterpart also observed for three out of seven QPE sources: \target, eRO-QPE2 \citep{Wevers2025}, and AT2019qiz \citep{Nicholl2024}.

We focus on two primary classes of models: `disk instabilities' and `orbiter/disk interactions'. These models are particularly relevant because their predictions are closely linked to the properties of the underlying accretion disk, for which our multi-wavelength analysis now provides the most precise estimates. However, it is important to acknowledge that all models attributing the origin of eruptions emission to a direct accretion phenomena (e.g., repeating partial TDE, but also disk instabilities) may already face challenges from the observed anti-clockwise `hysteresis loops' in the intensity (luminosity) versus hardness (temperature) parameter space during eruptions \citep[e.g., ][]{Arcodia2022,Vurm2024,Chakraborty2024}. This aspect, however, will not be addressed further here, as the timing properties of the eruptions are not the subject of this work. Still we aim to discuss some instabilities models, with emphasis on those for which direct predictions for the time scales and the (in)stability criteria are available, and can be confronted to the our data-inferred disk parameters.

When directly comparing the properties of the disk in the two epochs, it is crucial to first note that, although the disk's X-ray emission in 2014 was both brighter and harder (hotter) than the disk emission in 2018, it was neither brighter nor harder than the spectra of the 2018 eruptions at their peaks, as detailed in Appendix \ref{app:reduction}. Hence the absence of QPEs in 2014 cannot be attributed to a simple `detectability challenge', where the eruptions are present but are outshone by the disk. Instead, the lack of QPEs in 2014 is most likely intrinsic \citep[in agreement with ][ see their Appendix C]{Miniutti2023_noqpe}, meaning the eruption were either intrinsically absent or at least intrinsically much weaker.

\subsubsection{Disk Instabilities}\label{sec:inst}

The presence and recurrence time of disk pressure instabilities are determined by factors such as radiation pressure, magnetic field strength \citep{Kaur2023}, as well as the accretion rate and the black hole mass. For a classical non-magnetized disk, Equation 34 from \citet{Grzdzielski2017} \citep[or equation 13 of][]{Kaur2023} can be used to interpolate the recurrence time for radiation pressure instabilities, assuming an eruption amplitude relative to quiescence of $\gtrsim 10$ \citep{Miniutti2019}. This yields an expected recurrence time between 4 and 30 years, which is grossly inconsistent with observations.

However, \citet{Kaur2023} argue that instability timescales can be significantly shorter if the disk is magnetized. In this context, the criteria for instabilities depend not only on the properties of the disk and black hole (\Mbh, $\lambda_{\rm Edd}$ and the $\alpha$ parametrization of viscosity) but also on the dimensionless magnetic pressure scaling parameter ($p_0$).

The condition for instability to occur in a magnetized disk is that the Eddington ratio ($\lambda_{\rm Edd}$) must exceed the minimum Eddington rate ($\lambda_{\rm min, Edd}$), given by \citet{Kaur2023}'s Equation 10:
\begin{equation}\label{eq:edd_min}
    \lambda_{\rm min , Edd} = 0.066 \sqrt{p_0} M_6^{-1/16} \alpha_{-1}^{-1/16},
\end{equation}
where $M_6 = M_{\rm BH}/10^6$ \msun\ and $\alpha_{-1} = \alpha/0.1$. Below this threshold, the disk is stable. Therefore,  the condition for the QPE present in 2018 to be produce by this instability is $\lambda_{\rm Edd}^{2018} \geq \lambda_{\rm min, Edd}^{2018}$. If that is the case, the QPE recurrence time ($T_{\rm QPE}$) is given, in relation to the disk/MBH properties, as well $p_{0}$ and $\alpha$, by their Equation 14:

\begin{equation}\label{eq:tqpe}
    T_{\rm QPE} \approx 1.1\, {\rm yr} \,  \frac{M_6^{44/37} \lambda_{\rm Edd}^{38/37}}{p_0^{56/37}\alpha_{-1}^{30/37}}.
\end{equation}

Therefore, for a given $\alpha$, $p_0$ can be computed by:
\begin{equation}\label{eq:p_0}
 p_0 \approx  595 \frac{M_6^{44/56} \lambda_{\rm Edd}^{38/56}}{\alpha_{-1}^{30/56}} \left ( \frac{T_{\rm QPE}} {1 {\rm hr}} \right )^{-37/56}.
\end{equation}

\begin{figure}[h!]
	\centering
	\includegraphics[width=\columnwidth]{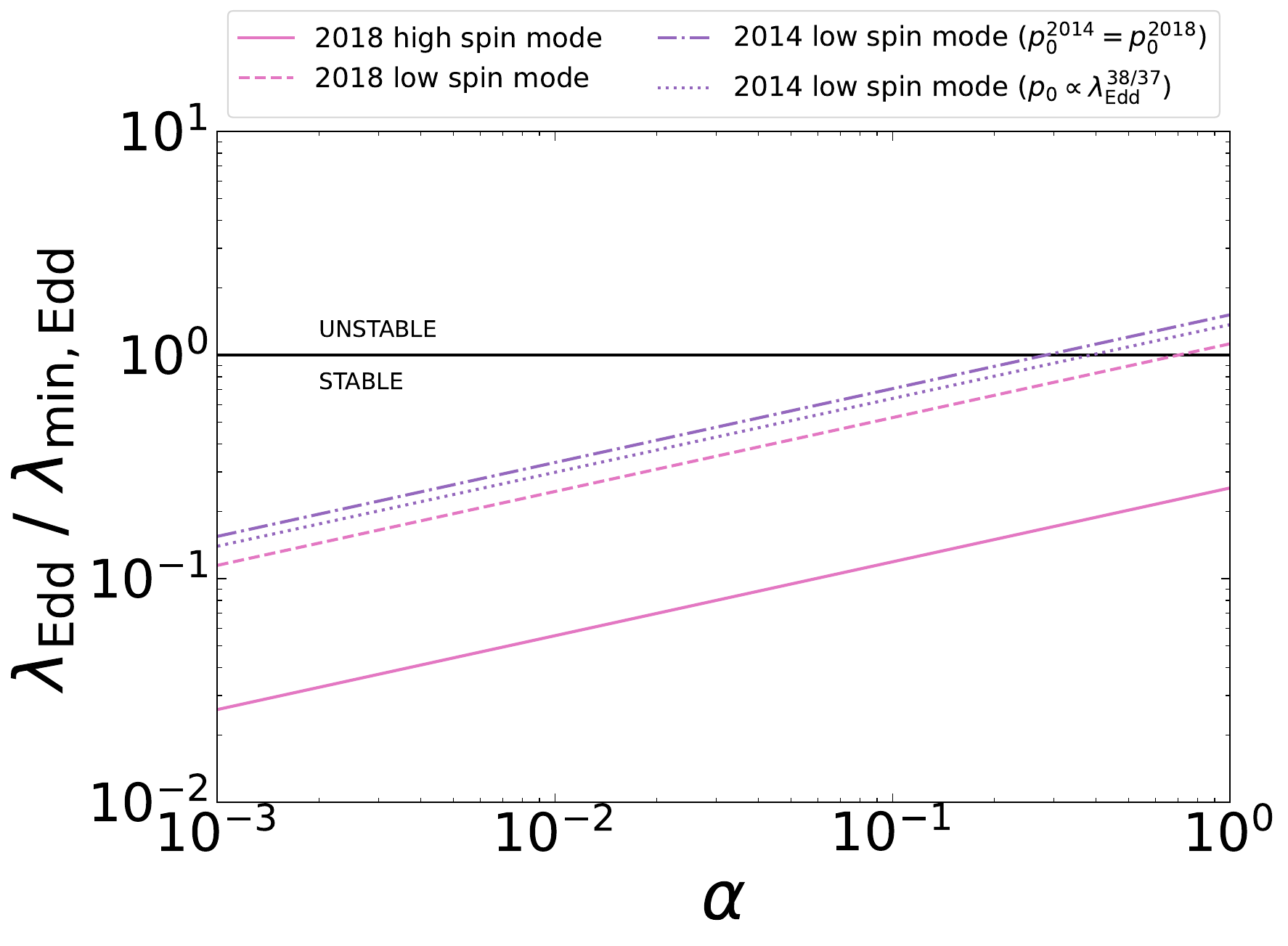}
 
	\caption{Instability condition for instability in \citet{Kaur2023}'s magnetized disks. The plot show the ratio between the measured Eddington ratios ($\lambda_{\rm EDD}$, Table~\ref{tab:par}), and the minimum  $\lambda_{\rm min, Edd}$, to trigger the instability, as a function of the $\alpha$ parametrization of the viscosity. No value of $\alpha$ (and $p_0$) can make the disk instable in 2018, while maintaining it stable in 2014 (see text for details).}
    \label{fig:kaur}
\end{figure}

The results from SED fitting in \S\ref{sec:kerrSED} (\Mbh and $\lambda_{\rm Edd}$) can now be applied to test whether the instability criteria are satisfied. Assuming the same range of $\alpha$ values as in \citet{Kaur2023}, Fig.~\ref{fig:kaur} displays $\lambda_{\rm Edd}/\lambda_{\rm min, Edd}$ in pink. Here, $\lambda_{\rm min, Edd}$ is calculated using Eq.~\ref{eq:edd_min}, with $p_0$ derived from Eq.~\ref{eq:p_0} and $T_{\rm QPE} = 9\,{\rm hr}$. The computation assumes the 2018 MBH/disk properties (as given in Table~\ref{tab:par}) for both the high-spin (solid line) and low-spin (dashed line) configurations.

For the high-spin scenario, it is evident that no value of $\alpha$ renders the disk unstable in 2018. However, in the low-spin solutions, the disk could, in principle, become unstable under this framework if $\alpha > 0.7$. Importantly, we also have disk properties from 2014, when $\lambda_{\rm Edd}$ was higher by approximately $30\%$, yet the disk was stable, with no QPEs observed. To compute $\lambda_{\rm min, Edd}^{2014}$, we require a value for $p_0^{2014}$. Two approaches are in principle possible: we can either assume that $p_0$ remains unchanged with varying $\lambda_{\rm Edd}$, i.e., $p_0^{2014} = p_0^{2018}$, or assume it scales as in Eq.~\ref{eq:p_0}, such that $p_0^{2014} = p_0^{2018} \left( \frac{\lambda_{\rm Edd}^{2014}}{\lambda_{\rm Edd}^{2018}} \right)^{38/56}$. In Fig.~\ref{fig:kaur}, we plot $\lambda_{\rm Edd}^{2014}/\lambda_{\rm min, Edd}^{2014}$ for both cases, using dotted and dot-dashed purple lines, respectively. For either approach, the disk should also have been unstable in 2014 for $\alpha \geq 0.7$, contradicting observations of a stable disk during that time. This discrepancy poses significant challenges for the framework.

One way to stabilize a higher $\lambda_{\rm Edd}$ disk would be to have efficient advective cooling in the innermost regions (the X-ray emitting zones). However, such effects only dominate at $\lambda_{\rm Edd} \gtrsim 0.68$ \citep{Kaur2023}, whereas $\lambda_{\rm Edd}^{2014} < 0.68$.
Another option would be to `fine tune' $p_0^{2014}$ (for some $\alpha > 0.7$ value) so that $\lambda_{\rm Edd}^{2014} < \lambda_{\rm min, Edd}^{2014}$ is imposed. This, however, would imply that the QPE recurrence time, $T_{\rm QPE}$, is in general a time-dependent quantity, i.e., $T_{\rm QPE}(t)$, and approximately a linear function of $\lambda_{\rm Edd}(t)$, as per Eq.~\ref{eq:tqpe}. Yet, \citet{Miniutti2023_longterm} analyzed \target's X-ray light curves across different quiescent $\lambda_{\rm Edd}$ values (e.g., comparing their XMM5 and XMM6 observations), and while they have identified a correlation between $\lambda_{\rm Edd}$ and the intensity of the eruptions, they found no systematic variation in $T_{\rm QPE}$ with changing $\lambda_{\rm Edd}$. This challenges the viability of such fine tuning approach.

In summary, given the disk properties at the two epochs and the observed recurrence time, no combination of parameters can make the disk stable in 2014 and unstable in 2018. Comparing our assumptions and results on \target with \citet{Kaur2023}'s: (i) they assume \Mbh\ $= 4 \times 10^{5}$ \msun\ based on a specific \Mbh$-\sigma_*$ relation \citep[following][]{Miniutti2019}, which is incompatible with the SED properties (\S\ref{sec:kerrSED}) for any combination of spin-inclination in the allowed parameter space (Fig.~\ref{fig:a_x_i}); (ii) they were able to recover a portion of the parameter space that makes 2018 disk unstable (similar to us), however, they only impose that the disk is unstable in 2018, but do not account for the requirement of stability in 2014, when the Eddington ratio was only $\sim$30\% higher. The simultaneous requirement for stability in 2014 and instability in 2018 is the key factor driving our distinct final conclusion compared to theirs.

The radiation pressure instability in an inner unstable disk zone, proposed by \citet{Pan2022, Pan2023} as the origin of QPEs, modifies the \citet{Sniegowska2023} model by removing the inner ADAF, better describing thermally dominated disk spectra. We can not analyze this model in detail, as the conditions distinguishing stable and unstable disks are unclear, making it uncertain why, in their framework, all (TDE) disks would not become unstable and show QPEs. Additionally, like \citet{Kaur2023}, \citet{Pan2022} do not address the fact that the disk while unstable in 2018, in 2014, at a just slightly higher accretion rate, was stable. Without explicitly stated instability criteria, the model currently lacks falsifiability and is not explored further. 

Of course, the above instability models are all based on formal instabilities which can be induced in accretion flows when the \cite{Shakura1973} $\alpha$-viscosity prescription is taken literally, and a steady-state disk is assumed. The fact that none of these models reproduce the observations could of course tell us that disk instabilities are not the cause of QPE flares, but this could equally simply be telling us that an $\alpha$-viscosity is an inadequate description of the complexities of magnetohydrodynamic turbulence, and that TDE disk cannot be treated by steady-state assumptions. Naturally, the data and analysis we have performed here cannot rule in/out any instability models which have not yet been invented, particularly those which are more realistic than a tuned $\alpha$-viscosity parametrization, and that are time-dependent in nature.

\subsubsection{Orbiter/Disk Interactions}\label{sec:emri}

For orbiter/disk collisions to occur \citep{Linial2023,Franchini2023,Tagawa2023}, the configuration requires that the radius of the orbiter ($R_{\rm orb}$) lies within the outer radius of the disk (i.e., $R_{\rm orb} \leq R_{\rm out}$). In the case of circular orbits, $R_{\rm orb}$ is related to the recurrence time of the QPEs ($T_{\rm QPE}$) by the relation $T_{\rm QPE} = \pi \sqrt{R_{\rm orb}^3 / G M_{\rm BH}}$. The probability distributions of $R_{\rm orb}$ are shown in Fig.~\ref{fig:radii_profile}.

Clearly, the fact that the \target disk was large enough to be intercepted by an orbiting body in 2018 is supportive of the general disk-orbiter collision paradigm. On the contrary, the fact that in 2014 the disk was {\it already} large enough to be impacted is difficult to reconcile in this framework. It is crucial to note that a disk more compact than $R_{\rm orb}$ at any of the analyzed epochs is entirely inconsistent with the observed UV spectral shape. Given the measured temperature, such a compact disk would exhibit a UV spectrum at least partially dominated by the Rayleigh-Jeans tail, whereas the observed spectra are dominated by the mid-frequency range shape. Moreover, if the disk were smaller than $R_{\rm orb}$ in 2014 but expanded to exceed $R_{\rm orb}$ by 2018, corresponding to a factor-of-few increase in size, a significant change in the UV spectral shape would be expected \citep[see Figure 1 and 2 in ][]{Guolo_Mummery2024}.

One possible interpretation of these results is that disk-orbiter collisions are not the origin of QPEs, and therefore the size of the disk in 2014 and 2018 is irrelevant to the onset (or not) of QPEs. This, trivially, would explain an apparent lack of correlation between disk size and QPE emission.  Alternatively, it is reasonable to expect \citep[and this is certainly the case within the framework of e.g.,][]{Linial2023} that more than just the physical extent of the disk determines whether or not QPE emission can be detected. One key part of the model of \cite{Linial2023} is that to reach the high measured temperatures of QPE flares ($kT \sim 100$ eV), the gas shock-heated in the collision must be photon-starved, as if it was in thermodynamic equilibrium it would be too cool ($kT \sim 10$ eV) to be detected. Photon starvation describes the state of a gas which is unable to produce enough photons to carry away the luminosity budget of the gas in thermodynamic equilibrium, and is an effect which occurs in supernovae \cite{Nakar10}, and even black hole accretion flows in extreme regimes \citep{Davis19, Mummery24Plunge}. As photon production by Bremsstrahlung scales with the density of the gas in the disk squared, it is possible that the collisions in 2014 would not have produced photon starved gas, and therefore the flares would be too cool to have been detected. Meaning, that the eruption should till be there, but their lower temperature when in thermodynamic equilibrium  ($kT \sim 10$ eV), would make the shocked gas to shine mostly in the unobserved extreme UV wavelengths, and therefore not detectable in the X-rays.
Then, as more material is lost through the inner edge of the disk, the disk density would drop to the point where in 2018 it is possible that the collisions produce gas which is photon starved, and the apparent temperature rises to X-ray detectable levels. 

While this may appear to be somewhat contrived, disk theory does predict a relatively quickly decaying disk density $\Sigma_{\rm disk} \sim M_{\rm disk}(t)/R_{\rm out}(t)^2 \sim t^{-15/16}$ for the \cite{Cannizzo1990, Mummery_Balbus2021} standard model. For reference, the disk cooling (which we observe) is predicted to follow a shallower decay $T_p(t) \sim t^{-19/64}$. Much more detailed analysis, using a full time-dependent disk theory must be performed to test this hypothesis rigorously. This will be the focus of future work.  

Finally, we note that \cite{lu23} pointed out another potential observational prediction of disk-orbiter  models. In \cite{lu23} the authors note that the stripping of material from an orbiting body, which is then added to the disk, would change the properties of the disk exterior to this point, as this outer region now behaves like a disk with a source of  angular momentum at its inner edge. Standard accretion solutions in this limit predict $\nu L_\nu \propto \nu^{12/7}$ \citep{Agol2000, Mummery_Balbus2021, lu23}. 

Although a disk model with a mid-frequency spectral shape of $\nu L_\nu \propto \nu^{12/7}$ — capable of joint X-ray/UV/optical fitting — has not been implemented in any X-ray fitting package, we can still disfavor this behavior for \target\ based on the available data. Specifically, the optical-UV spectral slopes in both 2014 (pre-QPEs) and 2018 (post-QPEs) remain unchanged and are well described by $\propto \nu^{4/3}$. With the standard $\propto \nu^{4/3}$ spectral shape, our recovered color excess $E(B-V)$ value is consistent with independent estimates derived from the Balmer decrement \citep{Wevers2024}. In contrast, adopting a steeper intrinsic shape ($\propto \nu^{12/7}$) would require a substantial increase in the intrinsic color excess to maintain the same observed spectral shape. This adjustment would result in a color excess from the SED fitting that is incompatible with the value derived from the Balmer decrement. Furthermore, the orbital radius at which this feeding would have to take place (to explain the recurrence time of the QPE flares) is well within the outer edge we measure from the SED (Fig.~\ref{fig:radii_profile}), meaning that the bulk of the disk would have to be in this state, and therefore this spectral shape change should be detectable with the onset of the QPEs sometime between 2014 and 2018. 

In the \cite{lu23} model the orbiting body transfers mass to the disk through Roche-Lobe overflow, but a collisional model may also add material to the disk through ram pressure induced stripping \citep{Linial2023,Yao2024}. While our observations do not rule out the collisional model, it does imply that the mass added from collisions must be significantly smaller than the local mass content of the disk at this radius. In particular the decaying temperature from discovery to around 2020 \citep{Miniutti2023_longterm} means that in this period the inner disk mass is decreasing and there was no significant mass replenishing, at least until the late-2020 rebrightening.

\section{Conclusions}\label{sec:conclusions}
In this paper, we reanalyze publicly available UV/optical imaging and UV spectroscopy data of the tidal disruption event (TDE) candidate and quasi-periodic eruption (QPE) source \target, alongside performing self-consistent joint X-ray and UV spectra analysis using state-of-the-art disk models across two epochs. Our conclusions are as follows:

\begin{itemize}
    \item Using profile decomposition, we demonstrate that a combination of a point-source and diffuse stellar component well describes the innermost region of \target's FUV and optical images (see Fig.~\ref{fig:profile_fit} and \ref{fig:point_source}). The point source dominates the inner ($0.5\arcsec \times 0.5\arcsec$) region's FUV flux, with the diffuse stellar contribution accounting for less than 5\%. The diffuse component's color is consistent with a moderately young stellar population (100--300 Myr). The high FUV intrinsic luminosity ($L_{\rm FUV} \gtrsim 10^{42}$~\ergs) of the point-source precludes attributing any significant fraction of its flux to a nuclear stellar cluster (NSC), as this would require unphysically high star-formation surface densities (see \S\ref{sec:image}).
    \item Analysis of the two epochs of UV spectra (1150--3150~\AA) reveals that, after correcting for Galactic and intrinsic extinction/attenuation, the spectra are well described, to leading order, by $\nu L_\nu \propto \nu^{4/3}$, characteristic of the mid-frequency range of a standard thin disk. We also find that the integrated FUV flux decreased by $\sim$10\% from 2014 to 2018 (see Fig.~\ref{fig:spec}).
    \item We self-consistently model the X-ray and UV spectra together with the photometry of the diffuse stellar population. Our results demonstrate that a relatively simple disk model, with three free parameters, can simultaneously account for the X-ray emission and most (aside from the minor stellar contribution) of the UV spectra (see Fig.~\ref{fig:SED}).
    \item Using a more realistic disk model, which incorporates relativistic corrections via numerical ray tracing, we constrain the disk inclination ($i$) to $31^\circ \lesssim i \lesssim 63^\circ$ (99\% posterior). The black hole spin ($a$) remains ambiguous, with two equally likely modes in the posterior (thought not value of the parameter space can be excluded): (i) a highly spinning black hole ($a = 0.95^{+0.02}_{-0.05}$) with $M_{\rm BH} = 7.5^{+1.0}_{-1.5} \times 10^{6}~M_\odot$ and $i = 59^{+4}_{-2}$ deg; or (ii) a slowly spinning black hole ($a = -0.19^{+0.48}_{-0.40}$) with $M_{\rm BH} = 1.5^{+0.5}_{-0.3} \times 10^{6}~M_\odot$ and $i = 41^{+7}_{-6}$ deg (see Fig.~\ref{fig:a_x_i} and \ref{fig:cool_expansion}). These bimodal posterior distribution, maybe be broken, by e.g., adding more X-ray spectra into the fitting.
    \item In 2014, the outer disk radius ($R_{\rm out}$) was $\mathcal{O}(10^3 R_{\rm g})$ for both solutions, a plausible size for a TDE disk if observed many years after disruption, though a fully time-dependent modeling of the source is still warranted. However, the evidence for simultaneous cooling ($T_{\rm p}^{2018} < T_{\rm p}^{2014}$) and expansion  ($R_{\rm out}^{2018} > R_{\rm out}^{2014}$) of the disk are overwhelming (see Fig. \ref{fig:cool_expansion}, \ref{fig:radii_profile}). This behavior points to the fact that \target's disk is not in a steady-state, and inferred evolution is (unique) characteristic of a flow that is fed close to the black hole and expands viscously with a decreasing total inner disk mass; therefore, strong evidence for a TDE origin for the disk (see \ref{sec:gsn_tde}).
    \item Using the inferred system properties from both epochs (\Mbh, $\lambda_{\rm Edd}$, and \Rout), we test QPE model predictions. We argue that no published disk instability model can simultaneously produce an unstable disk (with the observed recurrence time) in 2018 while making the disk stable (no QPEs) in 2014, given $\lambda_{\rm Edd}^{2018}/\lambda_{\rm Edd}^{2014}$ is only $\approx 0.7$ (see \S\ref{sec:inst}).
    \item For the disk/orbiter interactions models, although the large enough disk in 2018 is supportive of orbit disk collisions, the lack of the QPEs in 2014 when \Rout was already $\gg R_{\rm orb}$ ($= 30-200 R_{\rm g}$), is hard to be reconciled with the model, unless a very fine-tune $\Sigma_{\rm disk}$ evolution has occurred (see \S\ref{sec:emri}). This possibility can be rigorously tested in future work via fully time-dependent modeling of the flow.
    \item In general, our work highlights that when multi-epoch high-quality multi-wavelength data is available and is physically and self-consistently analyzed, such analysis allows for rigorous testing of QPE models, which currently struggle to reproduce all the observables simultaneously.
\end{itemize}

\textit{Acknowledgements} -- 
MG is grateful to G. Miniutti and other participants of the ``Galactic and Extragalactic X-ray Transients" conference, for insightful discussions on \target, and to D. Law for providing star formation rate surface density measurements from the MaNGA survey. MG is supported in part by NASA XMM-Newton grant 80NSSC24K1885.
This work was supported by a Leverhulme Trust International Professorship grant [number LIP-202-014]. MN is supported by the European Research Council (ERC) under the European Union’s Horizon 2020 research and innovation programme (grant agreement No.~948381) and by UK Space Agency Grant No.~ST/Y000692/1. The HST data presented in this article were obtained from the Mikulski Archive for Space Telescopes (MAST) at the Space Telescope Science Institute. The specific observations analyzed can be accessed via \dataset[doi:10.17909/xn57-rw37]{http://dx.doi.org/10.17909/xn57-rw37}. STScI is operated by the Association of Universities for Research in Astronomy, Inc. under NASA contract NAS 5-26555.

\bibliography{tde}{}
\bibliographystyle{aasjournal}

\appendix

\section{Data Reduction}\label{app:reduction}

\begin{figure}[h!]

	\centering
	\includegraphics[width=0.8\columnwidth]{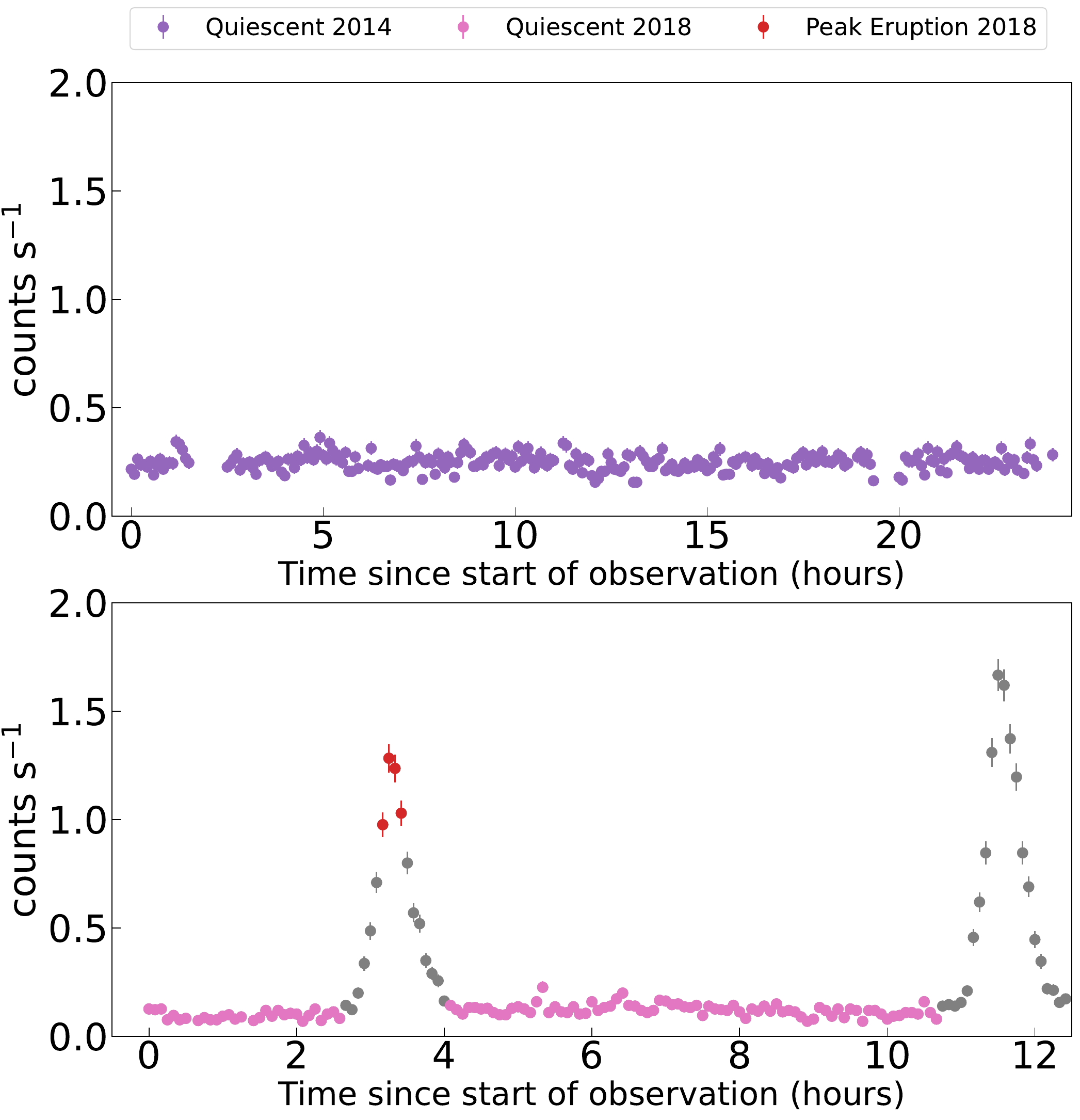}
    \caption{X-ray light curves of the \xmm observations, with GTIs stacked to create the spectra colored accordingly.} 
    \label{fig:lc_reduction}
\end{figure}

\textbf{X-ray:} The X-ray data set underlying this work is based on \xmm observations,  these observations were taken as part of Guest Observer (GO) programs (P.I. Miniutti, OBS-ID 0740960101 and 0823680101) and are public available in the \xmm archive. The observations were taken in Full Frame mode with the thin filter using the European Photon Imaging Camera \citep[EPIC;][]{Struder2001}. The observation data files (ODFs) were reduced using the \xmm Standard Analysis Software \citep[SAS;][]{Gabriel_04}.
The raw data files were then processed using the \texttt{epproc} task. 
Since the pn instrument generally has better sensitivity than MOS1 and MOS2, we only analyze the pn data. 
Following the \xmm data analysis guide, to check for background activity and generate ``good time intervals'' (GTIs), we manually inspected the background light curves in the 10--12\,keV band. 
Using the \texttt{evselect} task, we only retained patterns that correspond to single and double events (\texttt{PATTERN<=4}). The source spectra were extracted using a source region of $r_{\rm src} = 35^{\prime\prime}$ around the peak of the emission. 
The background spectra were extracted from a $r_{\rm bkg} =
108^{\prime\prime}$ region located in the same CCD. The ARFs and RMF files
were created using the \texttt{arfgen} and \texttt{rmfgen} tasks,
respectively. For the 2018 observation, in which QPE are present we select only GTIs where QPEs are not present, as illustrated in Fig.~\ref{fig:lc_reduction}, such that the `quiescent' disk emission is the resulting spectrum; no QPEs were present in the 2014 data. The 2014 and 2018 quiescent spectra have respectively 7820 and 1080 counts.

\begin{figure}[h!]
    \label{fig:spec_reduction}
	\centering
	\includegraphics[width=0.82\columnwidth]{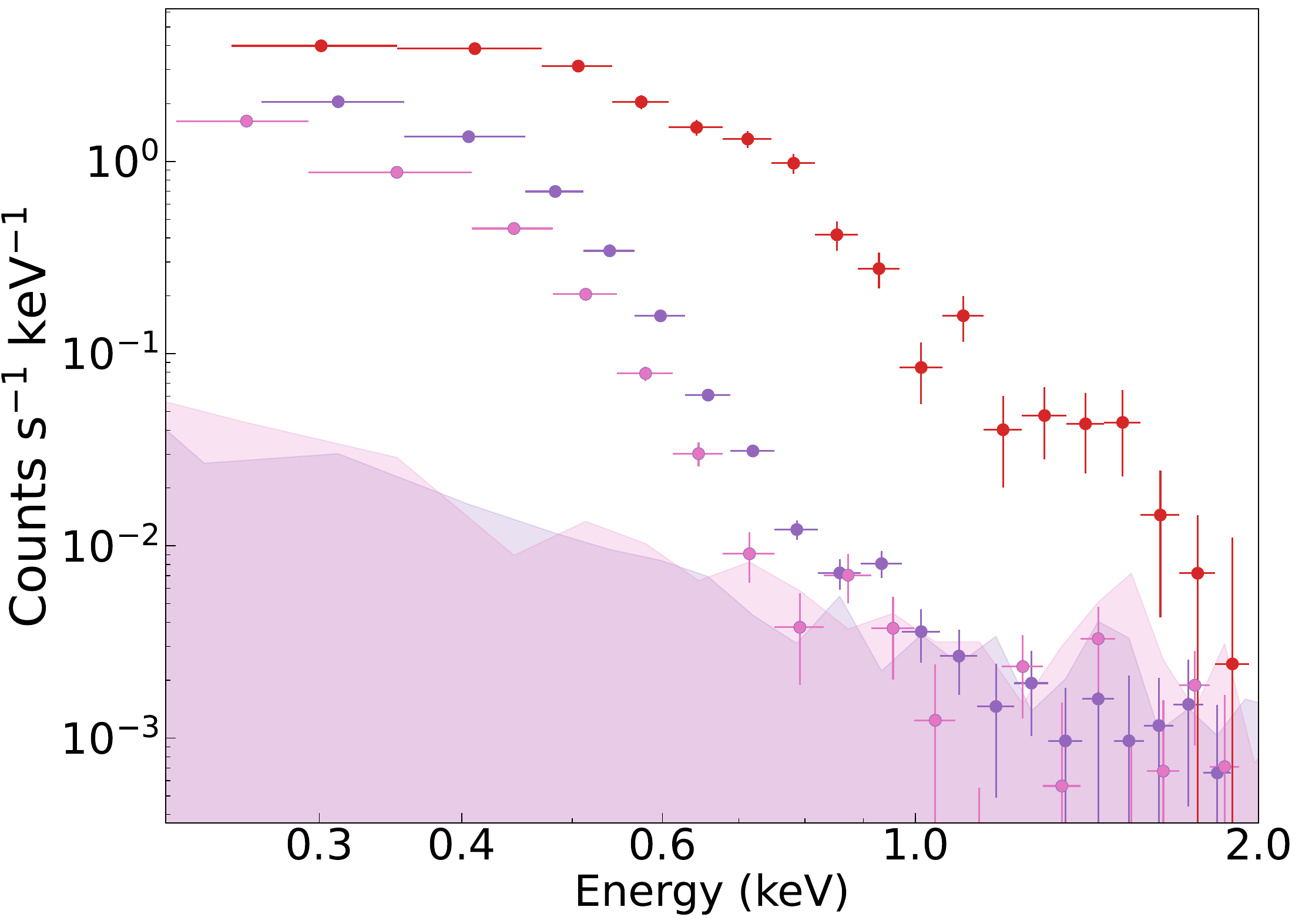}
    \caption{Background subtracted spectra produced from the observations (and GTI's) as shown Fig.~\ref{fig:lc_reduction}. Filled pink and purple regions, represent the background criteria used to determined the fittable range of the quiescent spectra. The eruptions peak in 2018 (red) were brighter and harder (hotter) than the quiescent emission in 2014 (purple). } 
\end{figure}

We also extracted a spectrum of one of the eruptions in the 2018 data, produced as described above, as shown in \ref{fig:spec_reduction}, it is clear that the spectra of peak of the eruption in 2018 were brighter (in every energy channel) than the quiescent emission in 2014, the spectrum is also harder. Therefore, the lack of QPE in 2014, must be intrinsic, in 2014 the QPE were either absent or were intrinsically much weaker. This interpretation is in agreement with \citet{Miniutti2023_noqpe}, in particular with their detectability simulations in their Appendix C.

\textbf{UV/optical Imaging:} Archival HST images of GSN 069 were retrieved from the Mikulski Archive for 
Space Telescopes (MAST). These  images were originally obtained in HST Cycle 27 (Program ID 16062; PI: 
Miniutti) in 2020 August and November.  We use three of the available images which were obtained with the 
Advanced Camera for Surveys (ACS): two far-ultraviolet (UV) band (F140LP, F165LP) images, and a wide-band 
optical image (F606W). We use the pipeline produced data products that have been corrected for the effects of charge transfer efficiency. For each image, the pixel counts in units of ${\rm electrons^{-1}}$ were converted to flux density
in units of \fluxdens using the \texttt{PHOTFLAM} keyword in the image headers.

\textbf{UV Spectra:} We start our analysis with the pipeline-reduced {\tt flt} files of visits {\tt ocm3} and {\tt odrx}. These include observations with the FUV-MAMA detector (G140L grating) and the NUV-MAMA detector (G230L grating), taken in two epochs (2014 and 2018). All observations employ the 52X0.5 slit. The relative orientation of the spacecraft V3 axis differs by 11.4 degrees between these two visits. To minimize host galaxy contamination, we perform a custom extraction using the {\tt x1d} task in the {\tt STIStools} package. Our aim is to obtain a nuclear spectrum within a 0.5\arcsec$\times$0.5\arcsec box. Because the encircled energy correction for STIS long-slit spectra is only available with certain extraction regions \citep{riley18}, we use a 21 pixel extraction region in the along-dispersion direction. This choice ensures that a point source encircled energy correction factor derived from standard star observations is available, and corresponds to 0.52\arcsec. Given the (small) difference in relative orientation of the dispersion axis, we then verify that neither slit orientation includes bright SF knots within the 0.5\arcsec extraction box that may artificially increase the resulting flux level. We use the custom co-add code\footnote{\href{https://spacetelescope.github.io/hst$\_$notebooks/notebooks/HASP/CoaddTutorial/CoaddTutorial.html}{Link to custom co-add code.}} made available through the Hubble Advanced Spectral Products \citep{hasp} to produce final co-added spectra for each epoch.

\section{Star Formation Rate Surface Density Estimates}\label{app:sfrsd}

The star formation rate surface density ($\Sigma_{\rm SFR}$) quantifies the rate at which stars are formed per unit area within a star-forming region. While the integrated star formation rate (SFR) is closely linked to the total amount of cold molecular gas available for conversion into stars \citep{Kennicutt1998}, $\Sigma_{\rm SFR}$ similarly correlates with the cold molecular gas surface density \citep{Roychowdhury2015, Pessa2021}. Consequently, $\Sigma_{\rm SFR}$ is a physically limited quantity constrained by the amount of cold gas present per unit area in a galaxy and the efficiency with which this gas can form stars. In the local universe, disk/spiral galaxies can have $\Sigma_{\rm SFR}$ spanning in a range of $-3.5 \lesssim {\rm log}(\Sigma_{\rm SFR}/M_{\odot} \, {\rm yr}^{-1} \, {\rm kpc}^{-2}) \lesssim 0$, as demonstrated by the $\Sigma_{\rm SFR}$ values derived from 1.4 million spaxels (spatially resolved regions within galaxies) across 4,517 galaxies in the MaNGA survey \citep{Barrera-Ballesteros2021,Law2022}. 

For a point source whose physical origin is attributed to stellar emission and whose physical size is constrained by the instrument's point-spread function (PSF), the star formation rate surface density ($\Sigma_{\rm SFR}$) can be estimated as:

\begin{equation}\label{eq:srfsd}
    \Sigma_{\rm SFR} = \frac{\rm \langle SFR \rangle}{\rm area} = \frac{\nu L_\nu}{\pi \, t_{\rm age} \, {\rm FWHM}^2} \left( \frac{M}{L} \right)_{\nu, t_{\rm age}}
\end{equation}

\noindent where $\nu L_\nu$ is the luminosity of the point source in a given filter of frequency $\nu$, ${\rm FWHM}$ is the full-width at half-maximum (in physical units) of the instrument, and $t_{\rm age}$ is the estimated age of the stellar population (which is reflected in, e.g., its color).

\begin{figure*}[b]
	\centering
    \includegraphics[width=0.9\columnwidth]{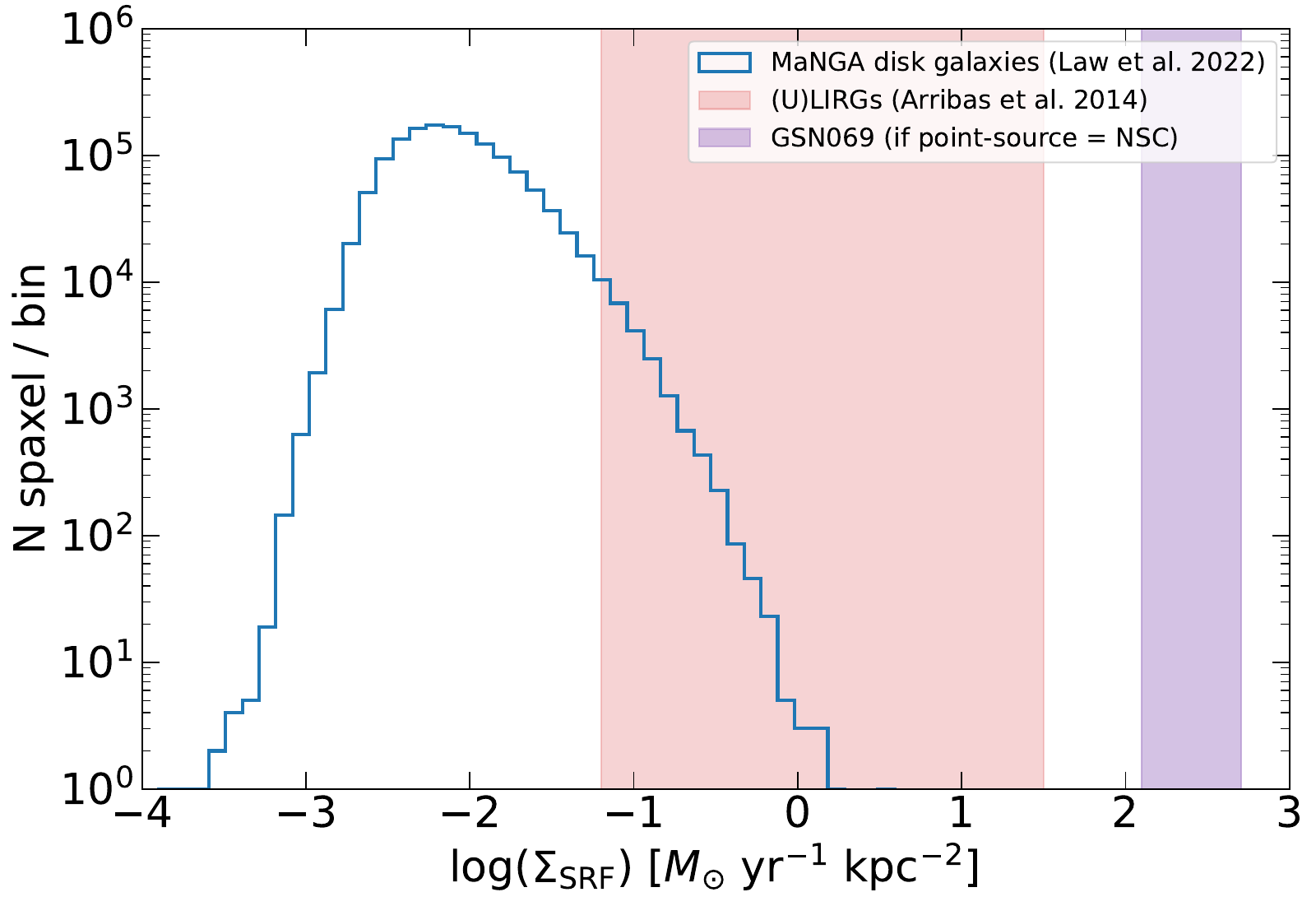}
    \includegraphics[width=0.9\columnwidth]{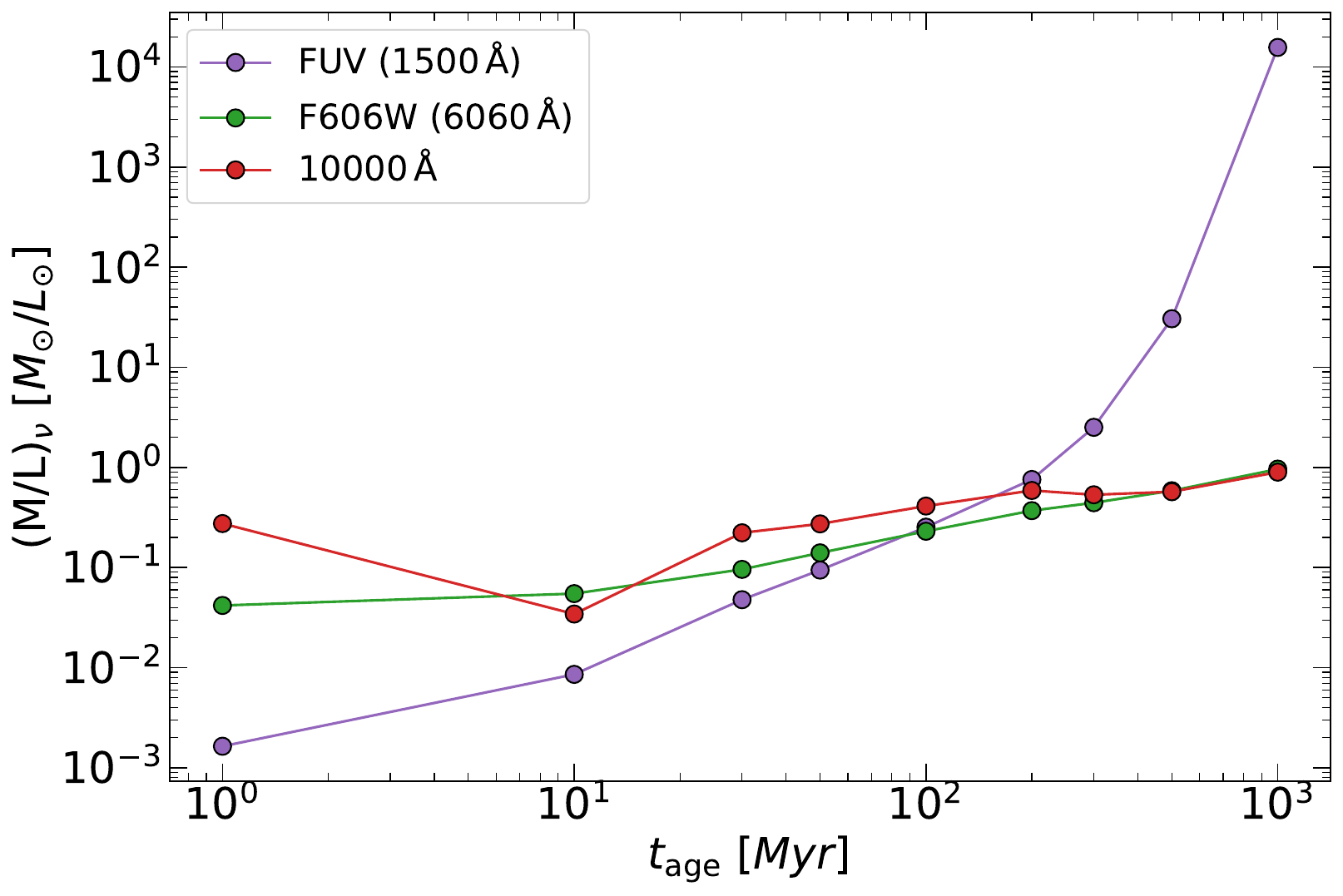}
    \caption{Left: Star formation rate surface density ($\Sigma_{\rm SFR}$) distributions. The blue histogram (and left y-axis) represents the distribution from the SDSS MaNGA survey, encompassing 1.4 million regions across $\sim4500$ star-forming disk galaxies \citep{Law2022}. The red band indicates the range of $\Sigma_{\rm SFR}$ observed in starburst and (ultra-)luminous infrared galaxies \citep{Arribas2014}. The purple contour illustrates the $\Sigma_{\rm SFR}$ that would result if the FUV-bright point source in \target\ were attributed to a nuclear stellar cluster. Right: Mass-to-light ratios derived from the simple stellar population models of \citet{Maraston05}.}
    \label{fig:SRFSD}
\end{figure*}

Lastly, $(M/L)$ is the mass-to-light ratio, which depends on both $\nu$ and $t_{\rm age}$, and is an output of any stellar population model \citep[e.g.,][]{Maraston05}. In the right panel of Fig. \ref{fig:SRFSD}, we plot $(M/L)$ for several wavelengths of interest, including the effective FUV filter (F140LP-F165LP) and F606W, as a function of $t_{\rm age}$.

By attributing the origin of the point source observed in \target to an unresolved nuclear stellar cluster (NSC), we can estimate the resulting $\Sigma_{\rm SFR}$ using Eq. \ref{eq:srfsd}. In this case, the FUV luminosity is $\nu L_\nu = 2 \pm 1 \times 10^{42}$ erg s$^{-1}$ ($\nu = 2 \times 10^{15}$ Hz), and the stellar population has an estimated age of $t_{\rm age} \approx 50$ Myr (see Fig. \ref{fig:point_source}). This corresponds to a mass-to-light ratio of ${\rm (M/L)} \sim 9.4 \times 10^{-2} \, M_{\odot}/L_{\odot}$ (see Fig. \ref{fig:SRFSD}), while the ${\rm FWHM}$ in physical units is approximately 35 pc. Substituting these values in Eq. \ref{eq:srfsd}, we obtain:

\begin{multline}\label{eq:gsn_srf}
    \tiny
    \Sigma_{\rm SFR} \approx 276 \, M_{\odot} \, {\rm yr}^{-1} \, {\rm kpc}^{-2} \times \\
    \left( \frac{\nu L_\nu}{2 \times 10^{42} \, {\rm erg \, s^{-1}}} \right) 
    \left( \frac{{\rm (M/L)}_{\nu, t_{\rm age}}}{0.094 \, M_{\odot}/L_{\odot}} \right) \times \\
    \left( \frac{t_{\rm age}}{50 \, {\rm Myr}} \right)^{-1} 
    \left( \frac{{\rm FWHM}}{35 \, {\rm pc}} \right)^{-2}.
\end{multline}

It is important to note that $\Sigma_{\rm SFR}$ is linearly dependent on $\nu L_\nu$ (Eq. \ref{eq:gsn_srf}). Hence, it is clear from the left panel of Fig.~\ref{fig:SRFSD} that attribution any relevant fraction of the FUV luminosity of \target's point source to a NSC would be highly contrived as this would mean a $\Sigma_{\rm SRF}$ unheard of for disk/spiral galaxies (\target is a disk/spiral galaxy), in the local universe. In fact, the inferred $\Sigma_{\rm SFR}$ would be higher then even the most intense starburst (ultra) luminous infrared galaxies \citep[ULIRGs,][]{Arribas2014} in the local universe; \target's host is clearly not a (U)LIRG, as although quite nearby, it is not even detected by the Infrared Astronomical Satellite \citep[IRAS,][]{Neugebauer1984}.

In summary, our analysis here, securely exclude the possibility that the FUV bright point-source that dominates the inner emission of \target, is produced, in any relevant fraction, by a young stellar population in a nuclear stellar cluster. Instead, as the relevant sections in the main text show, an accretion origin naturally follows.

\section{Supplementary Figures}\label{app:post}

\begin{figure*}
	\centering
	\includegraphics[width=0.9\textwidth]{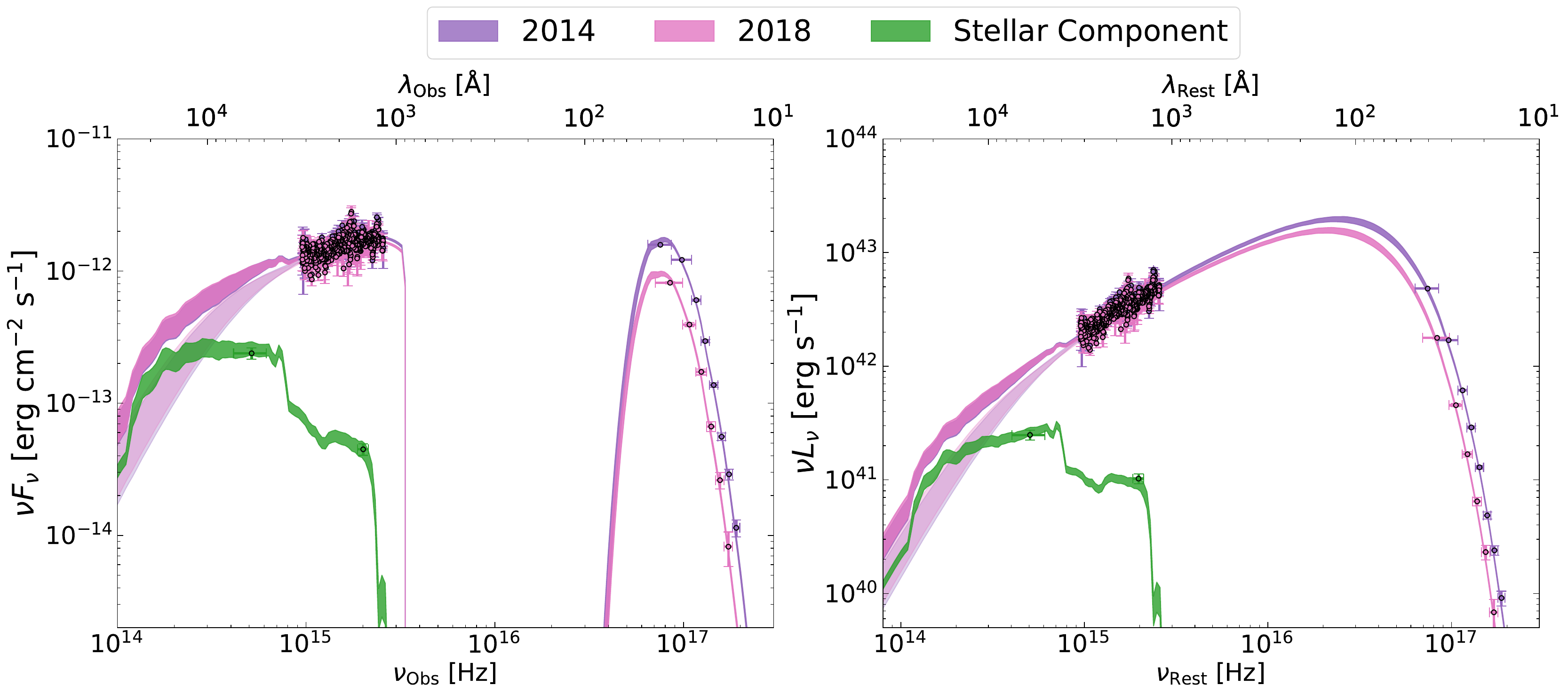}

	\caption{
    Results of the nested sampling fit of broad-band data. Purple colors refer to 2014 SED, pink to 2018, and green to the stellar component. For the two epochs the darker contours are total models (disk+stars), while the lighter contours are the disk-only emission. {\bf Top Left:} observed flux models (without any extinction/absorption correction) overlaid on the observed UV/optical data and the unfolded X-ray spectra. {\bf Top Right:} intrinsic luminosities (all absorption/extinction corrections), with the data points unfolded to the median values of the parameter posteriors.}
    \label{fig:kerrSED}
\end{figure*}

\begin{figure*}
	\centering
	\includegraphics[width=0.95\textwidth]{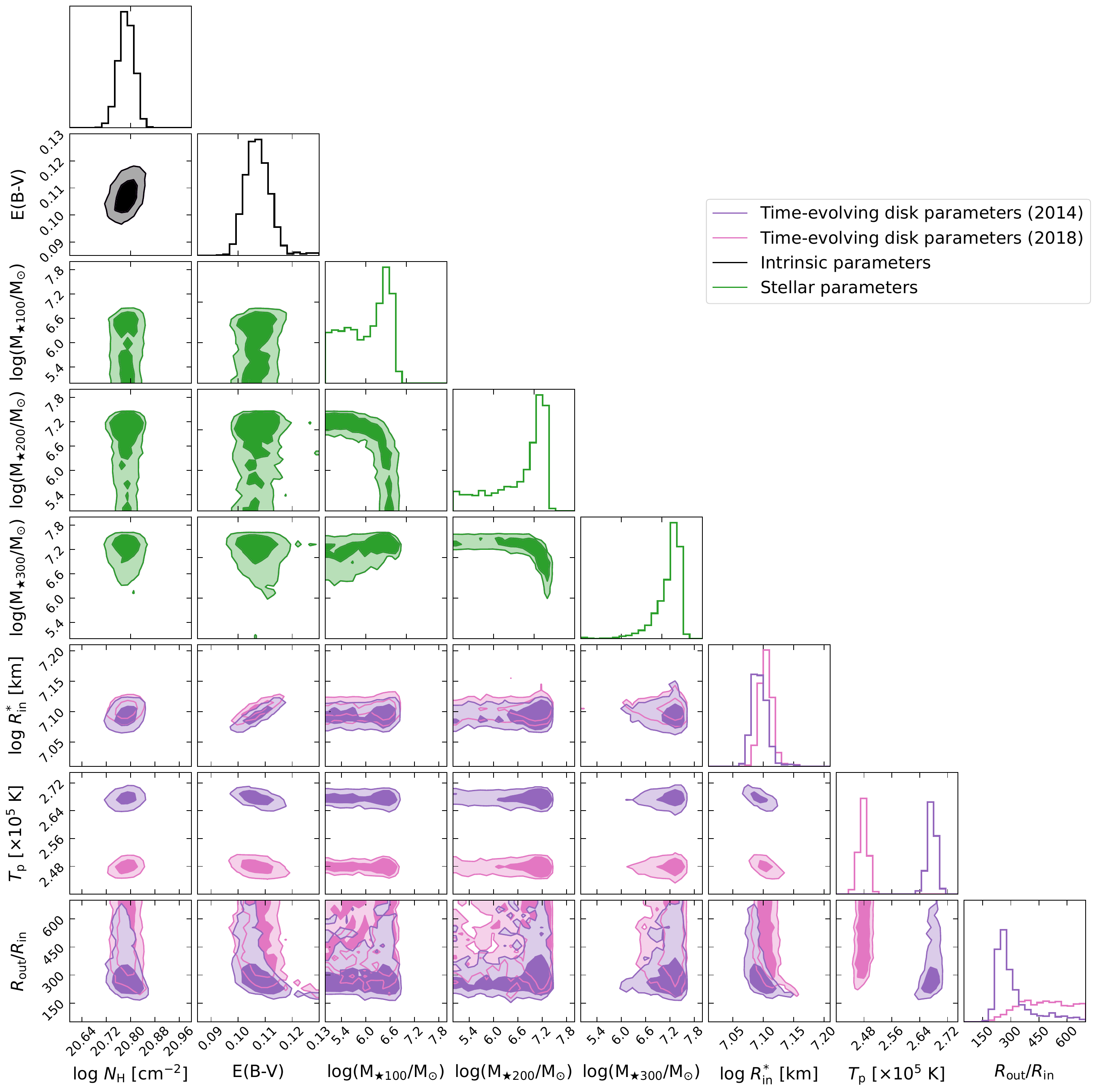}
	\caption{Posterior corner plot for the broad-band fit to the two epochs of \target, 2014 (purple), 2018 (pink), and stellar contribution (green), using \texttt{diskSED}. Intrinsic properties (fixed between epochs)  are shown in black. In the 2D histogram the contours shows 68\% and 95\% of the probability distribution.}
    \label{fig:post} 
\end{figure*}

\begin{figure*}[h]
	\centering
	\includegraphics[width=0.95\textwidth]{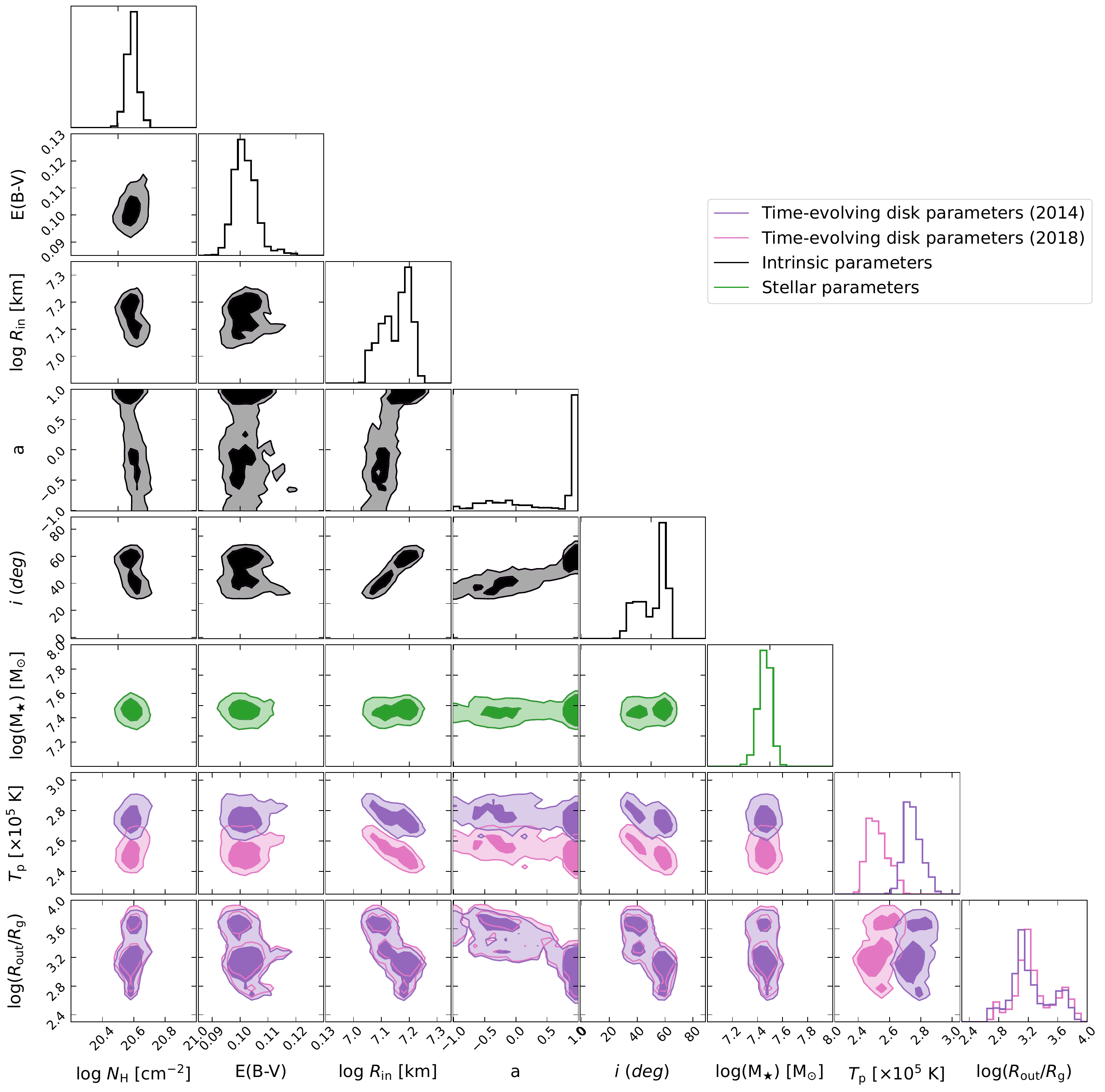}
	\caption{Posterior corner plot for the broad-band fit to the two epochs of \target, 2014 (purple), 2018 (pink), and stellar contribution (green), using \texttt{kerrSED}. Intrinsic properties (fixed between epochs) in black. In the 2D histogram the contours shows 68\% and 95\% of the probability distribution. The three stellar components masses were added, and only the total stellar mass ($M_\bigstar$) is displayed for clarity.}
    \label{fig:post_kerr}
\end{figure*}

\end{document}